\documentclass{amsart}
\RequirePackage{fix-cm}

\usepackage{wasysym}
\usepackage{amssymb,amsmath,stmaryrd,amsfonts}
\usepackage{mathtools,leftindex}
\usepackage{multirow}

\usepackage{array}

\usepackage{tikz,mwe,pifont}
\tikzset{boximg/.style={remember picture,red,thick,draw,inner sep=0pt,outer sep=0pt}}

\usepackage[numbers]{natbib}
\usepackage[margin=1.2in]{geometry}

\usepackage{enumitem}   

\usepackage{rotating}

\usepackage[colorlinks,citecolor=blue]{hyperref}

\begin{document}

\title[Small body dynamics]{Advanced Techniques in Stability Analysis of Trans-\linebreak Neptunian Objects}

\author[T.\,Kov\'acs]{Tam\'as Kov\'acs}
\address{
	Institute of Physics and Astronomy, E\"otv\"os University, Budapest, Hungary
}
\email{tamas.kovacs@ttk.elte.hu}

\date{\today}

\begin{abstract}
The trans-Neptunian region (30–50 AU) is a dynamically structured reservoir of icy planetesimals whose orbital architecture reflects resonant dynamics, chaotic transport, and long-term gravitational sculpting by the giant planets. This review synthesizes recent developments in the dynamical investigation of trans-Neptunian objects (TNOs), with an emphasis on mean-motion and secular resonances, as well as chaotic diffusion, in a system whose growing observational census makes it an ideal testbed for chaos detection methods. Classical indicators, including Lyapunov exponents, MEGNO, SALI/GALI, and frequency map analysis, provide the quantitative backbone for mapping TNO phase space and are complemented by modern approaches such as Lagrangian descriptors, the FAIR resonance identification method, entropy-based chaos indicators, and recurrence plot divergence methods. An anomalous diffusion framework, in which mean squared displacement scales as a power law in time, further enables classification of sub- and superdiffusive orbital transport. Machine learning has emerged as a powerful complement to traditional dynamical methods: surrogate classifiers, deep neural network solvers, and hybrid physics--data-driven frameworks together extend reliable prediction horizons in chaotic regimes and open new routes for Bayesian inference of migration scenarios. The review concludes that the most promising path forward lies in hybrid dynamical--statistical frameworks anchored to Hamiltonian dynamics, enabling efficient exploration of high-dimensional parameter spaces informed by the expanding body of trans-Neptunian observations.
\end{abstract}


\maketitle

\tableofcontents


\section{Introduction}
\label{sec:introduction}

Small bodies of the Solar System represent the most direct tracers of planetary formation and early dynamical evolution. Unlike the terrestrial planets and the gas giants, whose internal differentiation, atmospheric escape, and prolonged geological activity have erased much of their primordial record, trans-Neptunian objects (TNOs) preserve dynamical and compositional information from the protoplanetary disk \cite{Trujillo2001,Lykawka2012,dePra2022}. The Kuiper Belt, extending roughly from 30 to 50~AU, constitutes a dynamically structured reservoir of icy planetesimals that formed beyond the snow line and subsequently experienced gravitational sculpting by the giant planets. Since the discovery of (15760)~1992~QB$_1$ \cite{Jewitt1993}, the Kuiper Belt has evolved from a theoretical postulate into a cornerstone of modern planetary science, providing observational constraints on accretion, migration, resonance capture, and long-term chaotic transport \citep{Malhotra1995,Levison2003,Gladman2008,Emelyanenko2021}.

The present-day orbital architecture of the Kuiper Belt is highly non-uniform. Rather than forming a dynamically homogeneous disk, the belt is partitioned into multiple dynamical classes: cold classical objects with low inclinations and moderate eccentricities, hot classical objects with elevated inclinations, and resonant populations trapped in mean-motion commensurabilities with Neptune. There are also the scattered objects undergoing ongoing gravitational encounters and detached bodies whose perihelia lie beyond strong scattering influence \citep{Gladman2008}. This structural diversity encodes the combined effects of primordial disk conditions and subsequent planetary migration. In particular, the prominence of the 3:2 and 2:1 resonances with Neptune provided early dynamical evidence for large-scale outward migration of the ice giants \citep{Malhotra1995,Malhotra1996,Ida2000,Gomes2004,Hahn2005}.

In the framework of adiabatic theory \cite{Henrard1982a}, the probability of resonance capture during smooth migration depends on the drift rate of the resonant argument and the phase-space area enclosed by the separatrix \citep{Henrard1982b,Borderies1984,Cary1986,Tennyson1986}. Resonant capture occurs when the libration amplitude of the critical resonant angle remains bounded \cite{Nesvorny2009}. Migration-induced resonance sweeping excites eccentricities according to adiabatic invariance, producing correlations between semimajor axis and eccentricity that remain visible in resonant populations today~\mbox{\citep{Malhotra1995,Gomes2003}}.

Beyond isolated resonances, the Kuiper Belt occupies a region of phase space where resonance overlap and secular perturbations generate weak chaos. In the Hamiltonian formulation of the restricted three-body problem, the long-term dynamics may be described by a near-integrable Hamiltonian. When multiple resonances overlap, the Chirikov criterion~\cite{Chirikov1979} predicts the onset of global chaotic transport. In the outer Solar System, such overlap can occur between neighboring mean-motion resonances \cite{Wisdom1984} or between mean-motion and secular resonances, leading to slow diffusion in the semimajor axis and eccentricity \citep{Morbidelli2002}. This diffusive transport is a key mechanism linking the Kuiper Belt to the Centaur region and ultimately to Jupiter-family comets \citep{Duncan1997,Wiegert1999,Bonsor2012,Dones2015}.

Long-term stability in such weakly chaotic regimes cannot be inferred from short integrations alone. The largest Lyapunov exponent provides a local measure of exponential divergence \cite{Ott2002}, but in multi-resonant systems, short Lyapunov times do not necessarily imply rapid macroscopic diffusion. ``Sticky'' trajectories may linger near resonance islands for times vastly exceeding their Lyapunov timescales \cite{Milani1992}. Consequently, a hierarchy of diagnostic tools has been developed to quantify stability over gigayear timescales.

Spectral methods such as frequency map analysis track the drift of fundamental frequencies extracted from quasi-periodic decompositions of orbital elements \citep{Laskar1990,Laskar1993}. Regular trajectories exhibit constant fundamental frequencies, whereas chaotic orbits show measurable frequency diffusion. Variational indicators such as MEGNO \citep{Cincotta2000} accelerate convergence toward Lyapunov behavior and distinguish quasi-periodic and chaotic motion through the asymptotic value of a time-weighted divergence integral. Alignment indices (SALI, GALI) examine the evolution of deviation vectors and identify the dimensionality of invariant tori \citep{Skokos2001,Skokos2007}. Entropy-based approaches quantify phase-space exploration using information-theoretic measures derived from action-space partitioning \citep{Cincotta2003,Kovacs2022,Kovari2023}. Together, these methods provide complementary diagnostics for mapping the intricate stability landscape of the Kuiper Belt.

The large-scale structure of the belt further constrains models of planetary migration. The cold classical population, characterized by inclinations $i \lesssim 5^\circ$, displays a concentration near 44~AU known as the ``kernel'' \citep{Petit2011,Nesvorny2015}. Its narrow distribution in the semimajor axis and low dynamical excitation suggest partial in situ formation and limited perturbation. In contrast, the hot classical and scattered populations likely experienced strong scattering during a phase of giant-planet instability, as envisioned in the Nice model and its variants~\mbox{\citep{Tsiganis2005,Gomes2005,Levison2008}}. Numerical simulations show that temporary resonance trapping during Neptune's migration can implant objects into long-lived orbits while dynamically exciting their eccentricities and inclinations.

Detached objects with perihelia beyond $\sim$40~AU pose additional challenges. Proposed mechanisms for their origin include secular resonance sweeping, interactions with rogue planets, stellar flybys, or collective effects during migration \citep{Morbidelli2008}. These bodies highlight the sensitivity of the outer Solar System to both internal and external perturbations during its early history.


In recent years, advances in computational methods have expanded the scope of dynamical investigations. Large ensembles of N-body integrations, combined with sophisticated chaos indicators, allow systematic mapping of phase space across wide parameter ranges. Machine learning techniques have begun to complement traditional approaches by accelerating surrogate gravitational models and enabling automated classification of dynamical states 
 \citep{Tamayo2016,Pathak2018,Breen2020,Lemos2023}. Such tools facilitate exploration of high-dimensional parameter spaces relevant to migration scenarios and resonance capture probabilities.

The Kuiper Belt thus occupies a unique position in planetary science: it is simultaneously a fossil record of early accretion, a laboratory for nonlinear dynamical systems, and a bridge connecting the outer planetary region to cometary reservoirs. Its present architecture reflects the interplay of resonance dynamics, chaotic diffusion, collisional evolution, and planetary migration. By integrating analytical Hamiltonian theory, long-term numerical simulations, modern stability diagnostics, and emerging data-driven approaches, a coherent picture of Kuiper Belt evolution is gradually emerging.

This review aims to synthesize these developments. First, resonances and chaotic transport mechanisms shaping the belt are examined in Section~\ref{sec:dynamics}. Then, stability indicators and their application to trans-Neptunian dynamics are discussed (Section~\ref{sec:indicators}). Finally, modern computational and machine learning approaches that enhance our ability to interpret observational data and reconstruct the dynamical history of the outer Solar System are addressed in Section~\ref{sec:ml}. Through this synthesis, I seek to provide a comprehensive framework for understanding the long-term evolution and present structure of the Kuiper Belt.


\section{Resonances, Chaotic Transport, and Dynamical Structure}
\label{sec:dynamics}
\subsection{Mean-Motion Resonances}
\label{sec:mmr_in_kb}

Mean-motion resonances (MMRs) with Neptune constitute one of the most important dynamical sculpting mechanisms in the Kuiper Belt, governing not only the locations of long-lived populations but also their orbital excitation, stability boundaries, and transport pathways. A resonance arises when the ratio of the orbital period of a small body to that of Neptune satisfies an integer relation \cite{Murray1999}

\begin{eqnarray*}
      \frac{P}{P_N}=\frac{p}{q},\quad p,q\in \mathbb{Z}
\end{eqnarray*}
or 
equivalently
\begin{equation}
      pn-qn_N\approx 0,
\label{eq:mmr}      
\end{equation}
 where $n=\dot{\lambda}$ and $n_N=\dot{\lambda}_N$ are the respective mean motions. The resonant argument for the most common libration mode is given by
\begin{equation}
      \phi=p\lambda-q\lambda_N-(p-q)\varpi
\label{eq:res_angle}
\end{equation}
with $\lambda$ the mean longitude and $\varpi$ the longitude of the perihelion of the small body.
Objects for which $\phi$ librates rather than circulates are considered to be in resonance, as libration confines the orbital geometry and typically protects the object from close encounters with~\mbox{Neptune}.

The theory of MMRs in the Kuiper Belt is largely based on the averaged resonant Hamiltonian (following the canonical approach, e.g., \cite{Wisdom1980,Henrard1983,Morbidelli2002}). After averaging over fast angles, the resonant Hamiltonian reduces to a one-degree-of-freedom model of the form
\begin{equation*}
      \mathcal{H}_{\mathrm{res}}(J,\phi)=-\alpha J^2+\beta J +\epsilon e^{|p-q|}\cos\phi,
\end{equation*}
where $J\propto \sqrt{a}$ is the resonant canonical momentum, $\alpha,\,\beta$ are slowly varying frequencies, and $\epsilon$ scales with Neptune's mass and eccentricity. From this Hamiltonian, one can compute the resonant libration frequency
\begin{equation*}
      \omega_{\mathrm{lib}}\approx \sqrt{\alpha\epsilon}e^{|p-q|/2},
\end{equation*}
which determines both the resonance width and the adiabatic invariants relevant to capture during Neptune's migration \cite{Henrard1982b,Gomes1997}.

The resonance width in the semimajor axis, for low to moderate eccentricities, scales approximately as
\begin{equation}
  \Delta a \simeq a \left(\frac{M_N}{M_{\odot}}\right)^{1/2}e^{|p-q|/2},
  \label{eq:res_width}
\end{equation}
showing that high-order resonances become significantly wider at larger eccentricities~\mbox{\citep{Wisdom1982,Wisdom1985,Gladman1990,Holman1993,Malhotra1996}}. This explains why the 5:2, 7:4, and 9:4 resonances display non-negligible populations, despite the high order of these resonances $|p-q|,$ as demonstrated in numerical surveys \cite{Chiang2003b,Lykawka2005,Hahn2005,Forgacs2023}.

The dynamical significance of MMRs is strongly amplified by Neptune's outward migration. In the migration-sweeping scenario \cite{Malhotra1995,Hahn2005,Levison2008}, resonances move outward at\linebreak a~rate
\begin{equation*}
      \dot{a}_{\mathrm{res}}\approx \left(\frac{p}{p-q}\right)\dot{a}_N,
\end{equation*}
capturing objects when the adiabatic condition
\begin{equation*}
      \left|\frac{\dot{a}_N}{a_N}\right|<<\frac{\omega_{\mathrm{lib}}}{p}
\end{equation*}
is satisfied. Adiabatic capture preferentially affects low-inclination, low-eccentricity bodies, which are subsequently dynamically heated as continued migration within the 3:2 resonance excites their orbital elements. Once captured, an object's eccentricity grows according~to
\begin{equation*}
      \frac{\mathrm{d}e}{\mathrm{d}t}\approx -\frac{p-q}{p}\frac{\dot{a}_N}{2a}
\end{equation*}
a result verified in both analytic treatments and large-scale simulations of the Nice model~\mbox{\cite{Gomes2003,Levison2008}}.

One of the most remarkable outcomes of resonance sweeping is the 
''freezing in'' of eccentricities at the end of migration: when Neptune's migration slows or halts, resonant KBOs retain the large eccentricities acquired during the migration epoch. This mechanism naturally explains why the 3:2 resonance (Plutinos) contains many objects with $e\sim 0.2,$ an outcome originally noted by \cite{Malhotra1995} and later reproduced in population synthesis models~\mbox{\cite{Gomes2003,Nesvorny2000}}.

\begin{figure}[ht]
      \centering
      \includegraphics[width=0.85\textwidth]{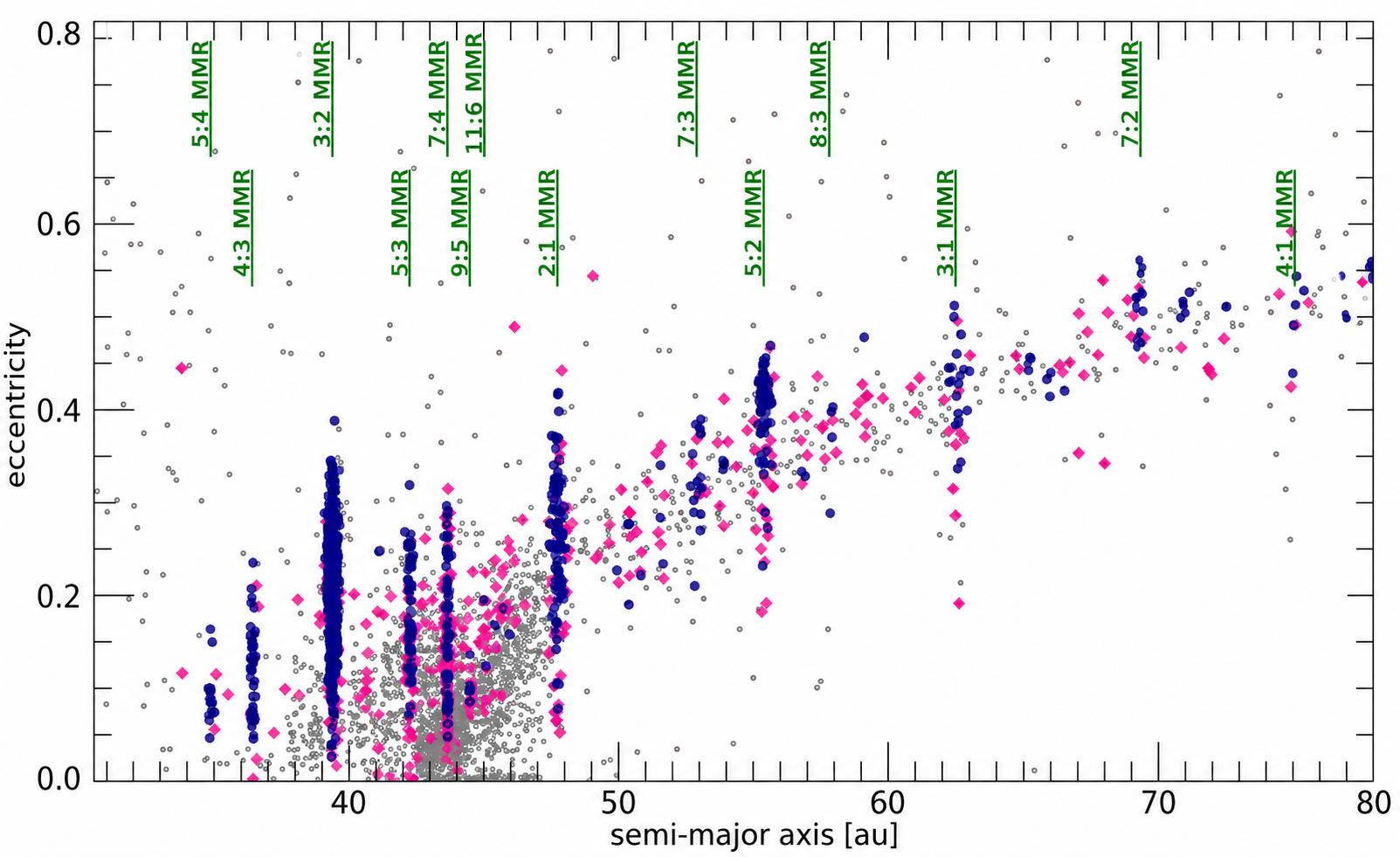}
      \caption{Distribution 
of Kuiper Belt objects in the semimajor axis--eccentricity space, illustrating the dominant mean-motion resonances with Neptune \cite{Forgacs2023}. Blue dots indicate long-term resonant objects, while pink dots show temporal libration of critical arguments. Gray dots represent the non-resonant TNOs. Green vertical lines are placed at the locations of major mean-motion commensurabilities, including the 3:2, 5:3, 7:4, 2:1, and 5:2 commensurabilities. Resonant confinement and resonance overlap structure play a central role in shaping the observed orbital distribution.}
      \label{fig:fde2024_mmr}      
\end{figure}

Beyond the well-known 3:2 and 2:1 resonances, the Kuiper Belt contains an intricate web of weak and high-order resonances that form dense clusters around \mbox{50--100~AU}~\mbox{\cite{Gladman2012,Lan2019,Forgacs2023}}; see Figure~\ref{fig:fde2024_mmr}. Laskar's \cite{Laskar1993} frequency map analysis reveals that dozens of resonances intersect, producing chaotic layers and resonance-sticking phenomena. It has been demonstrated \cite{Lykawka2007} that temporary resonance sticking is a significant mechanism for the long-term residence of scattering objects, with cumulative sticking times reaching millions of years.

High-order resonances often involve secondary resonances \cite{Malhotra1990,Henrard1992}, in which the libration frequency becomes commensurate with apsidal or nodal precession frequencies. Secondary resonances introduce fine chaotic structure inside major resonances—especially the 2:1 and 5:2 resonances—leading to chaotic diffusion in libration amplitude, eccentricity, and inclination. These chaotic layers are responsible for a significant fraction of the ''leakage'' of resonant objects into the scattering population \cite{Michtchenko1995}.

The interplay of eccentricity forcing and resonance phase protection means that resonant KBOs experience long-term stability even at perihelia approaching Neptune's orbit. The classical condition for phase protection, ensuring avoidance of close encounters, is
\begin{equation*}
      \Delta\phi_{\mathrm{lib}}<\phi_{\mathrm{crit}}(e,i),
\end{equation*}
where $\phi_{\mathrm{crit}}$ is the libration amplitude at which perihelion phases align unfavorably with Neptune’s longitude. Numerical experiments indicate stability up to high libration amplitudes, for instance, $\Delta\phi\approx 90^{\circ}$ for the 2:1, although these limits depend sensitively on inclination \cite{Volk2011,Li2014}.

Resonances also play a key role in shaping the dynamical boundaries of the classical belt. The observed ''edge'' near 47.8 AU corresponds to the inner branch of the 2:1 resonance, beyond which stable non-resonant orbits are sparse. This feature is a major observational confirmation of migration-sweeping dynamics \cite{Hahn2005}. The structure of the 2:1 resonance is particularly complex: its asymmetric libration centers split into “leading” and “trailing” islands \cite{Chiang2002}, whose population ratios provide constraints on the smoothness or stochasticity of Neptune's migration.

Transport across resonances can be quantified using analytic estimates of the chaotic diffusion coefficient
\begin{equation*}
      D_a=\frac{\langle(a(t)-a_0)^2\rangle}{2t},
\end{equation*}
which depends strongly on the degree of resonance overlap (e.g., in strongly overlapping regions such as between the 7:4, 12:7, and 5:3 resonances). Numerical studies show
\begin{equation*}
      D_a\sim 10^{-4}\;\text{to}\;10^{-3}\,\mathrm{AU}^2\,\mathrm{Myr}^{-1}
\end{equation*}
high enough to move bodies several tenths of an AU over Gyr timescales \cite{Tsiganis2005b,Shevchenko2010,Cincotta2018}.

Finally, MMRs provide a natural explanation for several key dynamical substructures in the Kuiper Belt:
\begin{itemize}
      \item The high-inclination “hot classicals,” produced by resonance capture and release \cite{Gomes2003,Gomes2005};
      \item The observed distribution of resonant populations, which encode Neptune's migration speed and jump magnitude \cite{Nesvorny2016};
      \item The long-term survival of the Plutinos, whose structure requires a migration timescale of 1--10 Myr \cite{Malhotra1995};
      \item The existence of distant resonant objects at 9:4, 7:3, and 5:2, which confirm that migration extended Neptune's semimajor axis by at least 7--10 AU ;
      \item The sharp boundaries of the cold classical belt, which are shaped by resonance passage without significant capture \cite{Levison2008}.
\end{itemize}

Mean-motion 
 resonances therefore act not only as dynamical sanctuaries for long-lived KBOs but also as signposts of the early Solar System's evolution, preserving a fossilized record of Neptune's dynamical instability, eccentricity damping, and outward migration.

\subsection{Secular Resonances and Long-Term Orbital Architecture}
\label{sec:sec_res_in_kb}

Secular resonances \cite{Murray1999} in the Kuiper Belt arise when the precession frequencies of a small body's longitude of perihelion $(g)$ or longitude of ascending node $(s)$ become commensurate with one of the fundamental eigenfrequencies of the giant planets, most importantly those associated with Neptune $(g_8,s_8).$ A general secular resonance condition is written as
\begin{equation*}
      g\approx g_8,\quad s\approx s_8,
\end{equation*}
though more complex combinations involving Jupiter and Saturn are possible. For a test particle at semimajor axis $a,$ the Laplace--Lagrange secular theory provides first-order estimates for the forced eccentricity and inclination arising from planetary perturbations. In this linear approximation, the disturbing function averaged over mean longitudes takes the form
\begin{equation*}
      \mathcal{R}_{\mathrm{sec}}=Ae^2+Bi^2+Cee_8\cos(\varpi-\varpi_8)+Dii_8\cos(\Omega-\Omega_8),
\end{equation*}
where coefficients $A,B,C,D$ depend on Laplace coefficients evaluated at the semimajor axis ratio $a/a_N$ \cite{Murray1999}. Although the linear theory captures broad trends, it underestimates the strongly nonlinear behavior occurring near resonances and in the presence of significant eccentricities, necessitating more detailed numerical or semi-analytic approaches \cite{Morbidelli2002,Michtchenko1995}.

One of the most important secular resonances in the Kuiper Belt is the $\nu_8$ resonance~\cite{Knezevic2022} defined by
\begin{equation*}
      g\approx g_8,
\end{equation*}
which primarily affects the eccentricity evolution of objects with semimajor axes near the boundary between the classical belt and the resonant populations. Numerical integrations~\mbox{\cite{Guzzo2002,Nagasawa2000}} have shown that $\nu_8$ can induce moderate eccentricity excitation, especially when combined with nearby mean-motion resonances such as the 5:3 and 7:4. The interaction between mean-motion and secular resonances often produces secondary resonances, where the libration frequency of an MMR becomes commensurate with the secular frequency mismatch (Equation~\eqref{eq:lib_secfreq}), producing a web of chaotic layers \cite{Morbidelli1995,Laskar1993}.

A related nodal resonance, the $\nu_{18}$ resonance, occurs when $s\approx s_8.$ This resonance can strongly excite orbital inclinations, particularly in the region $42<a<45$ AU. The existence of the ''hot classical'' population, characterized by large inclinations up to $i\sim 35^{\circ}$, may be partially explained by migration-driven passage of the $\nu_{18}$ secular resonance through the trans-Neptunian disk \cite{Gomes2003,Levison2008}.

Secular resonances play a particularly important role when combined with Neptune's orbital evolution during the planetary instability. According to the Nice model \cite{Desch2007,Crida2009}, the temporary excitation of Neptune's eccentricity substantially enhanced the secular forcing experienced by trans-Neptunian objects. The forced eccentricity vector \cite{Murray1999}
\begin{equation*}
      \mathbf{e}_{\mathrm{forced}}=e_f(\cos\varpi_8,\sin\varpi_8)
\end{equation*}
scales 
nearly linearly with Neptune's eccentricity and thus would have been much larger during the instability. As Neptune's eccentricity damped over $\sim$$10^6$ yr through interactions with the disk, the forced vectors shrank, leaving behind a residual distribution of free eccentricities consistent with the observed dichotomy between cold and hot classical~objects.

Secondary resonances play an especially important role in destabilizing objects near the 2:1 and 5:2 mean-motion resonances. A secondary resonance occurs when the libration frequency $\omega_{\mathrm{lib}}$ satisfies
\begin{equation}
      k\omega_{\mathrm{lib}}\approx|g-g_8|,\quad k\in\mathbb{Z}
      \label{eq:lib_secfreq}
\end{equation}
introducing fine chaotic structure within the resonance. Nesvorný \& Roig \cite{Nesvorny2001} demonstrated that secondary resonances embedded within the 2:1 mean-motion resonance can induce chaotic diffusion in eccentricity and inclination, explaining the wide range of orbital elements observed among Twotinos. The 5:2 resonance, one of the widest and most dynamically active resonances in the trans-Neptunian region, is permeated by a dense network of secondary resonances and secular commensurabilities, which shape its complex dynamical behavior 
\cite{Lecar2001}.

The combined influence of secular and mean-motion resonances can be visualized using frequency map analysis (FMA), as developed by Laskar \cite{Laskar1993,Laskar1999}. FMA computes the precession frequencies from numerical integrations and maps regions of constant frequencies, identifying diffusion zones where
\begin{equation*}
      |\Delta g|\sim 10^{-5}\,\text{yr}^{-1}
\end{equation*}
over tens of millions of years. These diffusion zones correspond to chaotic layers where particles evolve nonlinearly under overlapping resonant perturbations. FMA maps of the Kuiper Belt show strong frequency diffusion near the 7:4, 9:4, 12:7, and 5:3 resonances, consistent with resonance overlap theory \cite{Morbidelli2002}.

Long-term stability in the Kuiper Belt is therefore controlled by the delicate balance between resonant confinement and secular perturbations. In particular, secular effects shape the dynamical boundaries of the cold classical belt, which is characterized by low eccentricities and inclinations and is strongly depleted near regions influenced by the 
$\nu_8$ and $\nu_{18}$ resonances. Conversely, regions where mean-motion and secular resonances interact are highly chaotic, creating avenues for transport between the classical belt, resonant populations, and the scattering disk.

An important consequence of this complex secular structure is the existence of the observed ''kernel'' at $a\approx 44$AU \cite{Petit2011}, a narrow concentration of cold classical objects. Nesvorný \cite{Nesvorny2015} proposed that this feature arises from a discontinuity (''jump'') in Neptune's semimajor axis during the planetary instability, which shifted the locations of key resonances such as the 2:1 and $\nu_8.$ This resonance-sweeping interruption allowed a small region of the primordial Kuiper Belt to escape significant excitation, leaving a dynamically pristine subpopulation.

\begin{figure}[ht]
      \centering
      \includegraphics[width=0.75\textwidth]{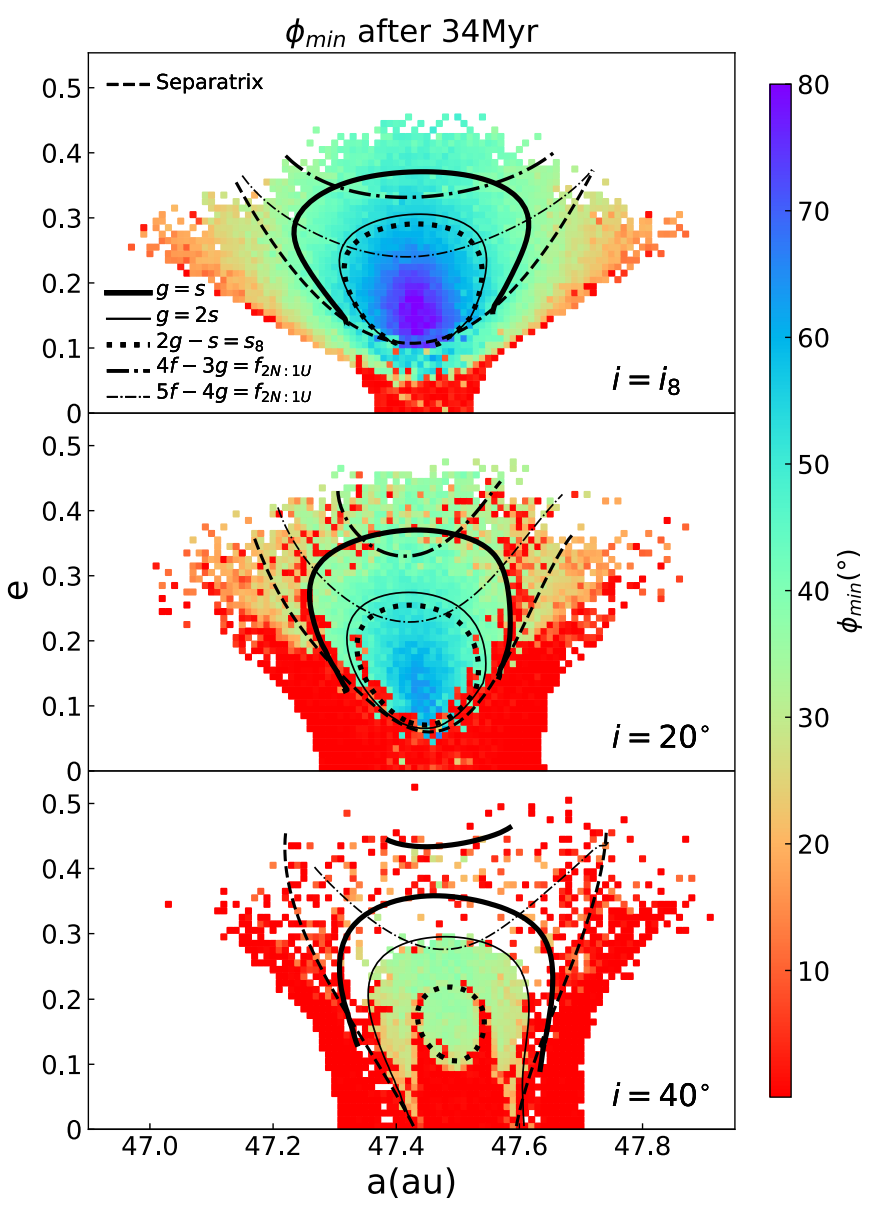}
      \caption{The 
minimal resonance angle $\phi_{\mathrm{min}}$ (related to Equation~(\ref{eq:res_angle})), an indicator of orbital stability, is color coded during the 34Myr integration time in the $(a, e)$ plane. The 1:2 MMR is explored according to the locations of the secular resonances for various inclinations, $i=0,\,20,\,40$~degs. For low inclinations (top and middle panels) the $g=2s$ and $s_8$ proper frequencies act almost identically; that is, they are responsible for stable motion. However, for large inclination (bottom panel), the Kozai mechanism dominates at large eccentricities and causes unstable dynamics. The dashed line separatrix indicates the border between the horseshoe and asymmetric resonances. Source of figure:~\cite{Li2024}.}
      \label{fig:li2024}
\end{figure}

Secular resonances also contribute to the shaping of detached KBOs. Particles with large perihelia ($q>40$ AU) can experience slow secular perturbations that lift their perihelia out of Neptune-crossing configurations. This mechanism can work with resonance sticking, Kozai cycles within high-order resonances (Figure~\ref{fig:li2024}), and external perturbations from passing stars or hypothetical distant planets \cite{Gomes2008,Batygin2016}.

Finally, secular architecture plays a central role in the chaotic transport of objects from the Kuiper Belt to the Centaur region. As a particle's eccentricity is secularly modulated, episodes of low perihelion distance may trigger close encounters with Neptune, leading to scattering. The timescale for secular modulation of perihelion can be estimated from the frequency spacing
\begin{equation*}
      \tau_{\mathrm{sec}}\sim \frac{2\pi}{|g-g_8|}
\end{equation*}
ranging from a few Myr to several tens of Myr, consistent with the timescales observed in long-term numerical integrations \cite{Duncan1995}.


\subsection{Proper Orbital Elements}
\label{sec:prop_elem_in_kb}

The concept of proper orbital elements \cite{Milani1990,Milani1992b} is fundamental for understanding the long-term architecture of the Kuiper Belt and for distinguishing intrinsic dynamical structure from transient perturbations. In analogy with the asteroid belt \cite{Morbidelli1991,Beauge2001}, proper elements represent quasi-invariant quantities that remain approximately constant on secular timescales, filtering out short-period oscillations due to planetary perturbations. For a trans-Neptunian object (TNO), the proper semimajor axis $a_p,$ proper eccentricity $e_p,$ and proper inclination $i_p$ are defined such that
\begin{equation*}
      a(t)=a_p+\delta a(t),\quad e(t)=e_p+\delta e(t),\quad i(t)=i_p+\delta i(t),\quad
\end{equation*}
where $\delta a,\,\delta e,$ and $\delta i$ encapsulate forced oscillations driven primarily by Neptune's gravity. The forced components arise from both secular perturbations \cite{Murray1999}, as described in \mbox{Section~\ref{sec:sec_res_in_kb}}, and proximity to mean-motion resonances. Proper elements provide a dynamical fingerprint of each object, allowing classification into dynamical families \cite{Lykawka2007b}, identification of clusters \cite{Benecchi2013,Vilenius2018,Marsset2019,Khain2020}, and constraints on the processes that shaped the outer Solar System~\mbox{\cite{Lykawka2009,Lykawka2011}}.

\begin{figure}[ht]
      \centering
      \includegraphics[width=0.85\textwidth]{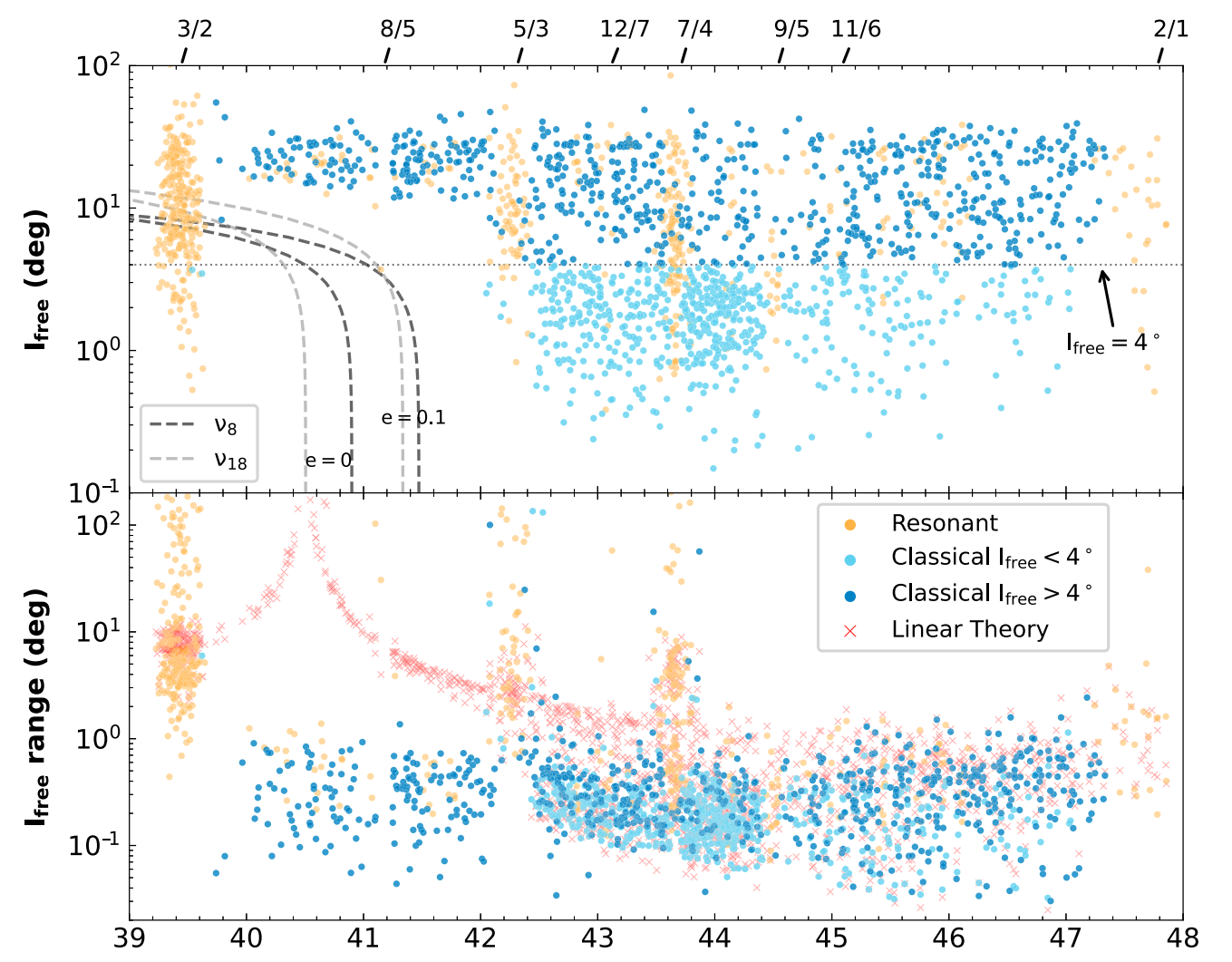}
      \caption{The 
 $i_{\mathrm{free}}$ values for 10Myr integration as a function of the barycentric semimajor axis. Top panel: The resonant (yellow) and classical (light ($i_{\mathrm{free}}<4^{\circ}$) and dark blue $i_{\mathrm{free}}>4^{\circ}$) TNOs are marked. The $\nu_8$ and $\nu_{18}$ secular resonances are also depicted for two eccentricity values. Bottom panel: $i_{\mathrm{free}}\;\text{range} = \max i_{\mathrm{free}} - \min i_{\mathrm{free}}$ indicates that the method of double-averaged Hamiltonian preserves inclination conservation, but the Laplace--Lagrange linear theory fails at $\nu_{18}$ secular resonance. In addition, neither method works well in the proximity of low-order mean-motion resonances (3:2, 5:3, and 7:4). Source of figure: \cite{Huang2022}.}
      \label{fig:huang2022} 
\end{figure}

Proper elements in the Kuiper Belt cannot be determined analytically using classical Laplace--Lagrange theory alone because many KBOs experience significant secular modulation, resonance interactions, and chaotic diffusion over timescales of tens of millions of years (see Figure~\ref{fig:huang2022}). As a result, proper elements are typically computed through numerical integrations using digital filtering or frequency analysis methods. Kne{\v{z}}evi{\'c} and collaborators pioneered such techniques for the asteroid belt, and similar methodologies have been adapted for the Kuiper Belt \cite{Knezevic2000,Nesvorny2000,Nesvorny2001}. The frequency map analysis (Section~\ref{sec:class_chaos_det}) method computes the dominant frequencies in the orbital evolution and extracts the free eccentricity $e_p$ and free inclination $i_p$ by decomposing the motion into forced and proper components:
\begin{eqnarray*}
      e(t)\exp(\iota\varpi(t))&=e_f\exp(\iota g_8t)+e_p\exp(\iota gt)\\
      i(t)\exp(\iota\Omega(t))&=i_f\exp(\iota s_8t)+i_p\exp(\iota st)\\
\end{eqnarray*}
where $(e_f,i_f)$ are the forced components tied to Neptune's precession frequencies $(g_8,s_8)$ and $(g,s)$ are the particle's proper precession frequencies, while $\iota$ denotes the imaginary~\mbox{unit}.

For dynamical classification, one of the most important quantities is the proper semimajor axis $a_p,$ which is typically much more stable than instantaneous $a(t)$ except for objects undergoing significant resonance interactions. In resonance-free regions, $a_p$ varies by less than $10^{-4}$ AU over gigayear timescales. However, near the 5:3, 7:4, and 2:1 resonances, objects can display oscillations \cite{Morbidelli2007} to
\begin{equation*}
      \delta a(t)\approx 0.1-0.3 \text{AU,}
\end{equation*}
necessitating long integrations, typically 50-200 Myr, to isolate the proper value. These long-term integrations are essential for distinguishing members of the cold classical belt~\cite{Li2014}, whose proper inclinations cluster around $i_p\lesssim 5^{\circ},$ from the high-inclination “hot classicals,” whose proper inclinations extend up to $i_p\sim 35^{\circ}.$ The distribution of proper elements clearly reveals that these two groups have distinct dynamical origins, likely reflecting differences between in situ formation and capture mechanisms during Neptune's migration \cite{Levison2008,Nesvorny2015b}.

A unique challenge arises when defining proper elements for resonant KBOs \cite{Gladman2012}, such as Plutinos \cite{Nesvorny2000a} in the 3:2 resonance and Twotinos in the 2:1 resonance. In these cases, the resonant argument $\phi$ librates and couples the evolution of $a,e,i$ through the resonant Hamiltonian. Proper elements for resonant bodies are not single constants but must be defined in the context of the resonant dynamics, typically through the invariants of the pendulum-like libration \cite{Henrard1982a}:
\begin{equation*}
      J_{\mathrm{res}}=\oint p_{\phi}\mathrm{d}\phi,      
\end{equation*}
where $p_{\phi}$ is the momentum conjugate to $\phi.$
The libration amplitude, resonant action, and libration center all function as generalized proper elements for resonant bodies \cite{Morbidelli2002}. Frequency analysis can still be applied, but the extracted proper frequencies are modified by the resonant dynamics and often include combinations of the libration frequency $\omega_{\mathrm{lib}}$ and the apsidal precession rates (see Equation~\eqref{eq:lib_secfreq}).

Proper-element computation becomes significantly more complicated in chaotic regions, where resonant overlap or secondary resonances induce frequency drift. In such regions, the diffusion coefficient
\begin{equation*}
      D=\frac{\langle(\delta a)^2\rangle}{2\Delta t}
\end{equation*}
can reach values of $10^{-4}-10^{-3}\text{AU}^2\text{Myr}^{-1}$ \cite{Tsiganis2005b}. The rapid diffusion implies that proper elements must be defined statistically, typically by identifying quasi-stable intervals lasting a few million years and averaging the orbital parameters within these intervals. This approach is commonly used for scattering objects and Centaurs, whose semimajor axes can vary by tens of AU on short timescales.

Proper elements serve as a key tool for identifying Kuiper Belt substructures. The ''kernel'' near 44 AU, for example, appears as a narrow concentration in proper semimajor axis, with
\begin{equation*}
  a_p\approx 44.0\pm 0.1\text{AU,}\quad i_p<3^{\circ},
\end{equation*}
suggesting that this cluster survived Neptune's migration with minimal perturbation. Its remarkably narrow proper-element distribution strongly supports the hypothesis that the kernel was formed by a discontinuous jump in Neptune's semimajor axis, decoupling this region from resonance sweeping \cite{Nesvorny2015}.

Proper-element analysis also helps identify collisional families. While no fully secure collisional family has been universally accepted in the Kuiper Belt, the Haumea family is widely regarded as the best candidate \cite{Brown2007}. Its members occupy a compact region in proper $a_p,e_p,i_p$ space, a distribution that is difficult to reproduce through dynamical means alone. Numerical reconstruction of the family requires precise determination of ejection velocities and dynamical diffusion in proper space, further demonstrating the utility of proper-element techniques.

Another important application involves comparing the dynamical temperature of different subpopulations. The cold classical belt has small proper values
\begin{equation*}
  e_p<0.05,\quad i_p<5^{\circ},
\end{equation*}
indicating that its current structure likely reflects primordial conditions rather than significant dynamical excitation. In contrast, resonant and scattering populations show large spreads in proper eccentricity and inclination, consistent with the excitation expected from resonance sweeping, scattering events, and chaotic transport during the early Solar System~\mbox{\cite{Gomes2003,Levison2008}.}

Finally, proper-element catalogs are essential for interpreting Kuiper Belt observational surveys, such as CFEPS \cite{Jones2006}, OSSOS \cite{Bannister2018}, and DES \cite{Elliot2005}. Because instantaneous orbital elements vary significantly with secular phase, population models must be constructed in proper-element space to avoid systematic biases. The OSSOS collaboration has demonstrated that synthetic populations with realistic proper-element distributions produce markedly better matches to the observed orbital distribution than those based on instantaneous osculating elements \cite{Bannister2016}.


\subsection{Resonance Overlap and Chaotic Transport}
\label{sec:transport_in_kb}

The dynamical structure of the Kuiper Belt is strongly shaped by the phenomenon of resonance overlap, a mechanism that generates large regions of chaotic motion and provides efficient pathways for long-range orbital transport. Resonance overlap occurs when the widths of neighboring mean-motion resonances (MMRs) or secondary resonances intersect in phase space, allowing a small body's orbit to wander stochastically between resonant islands. This mechanism is a general feature of systems governed by Hamiltonian perturbation theory, first formalized by Chirikov \cite{Chirikov1979}, and has been extensively applied to Solar System dynamics \cite{Wisdom1980,Laskar1993,Morbidelli2002}. In the Kuiper Belt, where the density of high-order resonances increases with semimajor axis, resonance overlap plays a central role in shaping the scattering population, the instability of the outer classical belt, and the structure of regions near the 2:1, 5:3, and 5:2 resonances.

For two nearby MMRs of order $|p-q|$ and $|p'-q'|$ resonance overlap occurs when the sum of their half-widths in the semimajor axis exceeds their separation:
\begin{equation*}
  \Delta a_{p:q}+\Delta a_{p':q'}\gtrsim |a_{p:q}-a_{p':q'}|.
\end{equation*}
The 
resonance width $\Delta a_{p:q}$ increases with eccentricity approximately as described by Equation~(\ref{eq:res_width}),
so that even high-order resonances (e.g., 7:4, 9:5, 12:7) become dynamically significant at moderate eccentricities ($e\gtrsim 0.1$). This scaling explains why the region between 42 and 48 AU is particularly prone to overlap-induced chaos: secular forcing from Neptune and transient excitation during migration drive KBOs to eccentricities where many resonances widen simultaneously.


The Hamiltonian structure of overlapping resonances can be described using the Chirikov diffusion model \cite{Cincotta2014}, in which the local chaotic diffusion coefficient in action space is
\begin{equation*}
  D_j\approx\frac{(\Delta J)^2}{2T_{\mathrm{hop}}}
\end{equation*}
where $\Delta J$ is the characteristic jump in the resonant action during a resonance crossing~\mbox{\cite{Nesvorny2015b,Lawler2019,Morbidelli2020}} and $T_{\mathrm{hop}}$ is the typical hopping time between resonances. Translating this into semimajor axis diffusion yields
\begin{equation*}
  D_a\simeq\frac{\langle(a(t)-a_0)^2\rangle}{2t}
\end{equation*}
with values in the Kuiper Belt ranging from
\begin{equation*}
  D_a\sim 10^{-6}-10^{-4}\text{AU}^{2}\text{Myr}^{-1}
\end{equation*}
depending on eccentricity and proximity to strong resonance chains \cite{Tsiganis2005b}.

One of the most striking consequences of resonance overlap is the creation of chaotic layers surrounding major resonances. In numerical surveys by Nesvorny \& Roig \cite{Nesvorny2000}, the 2:3 resonance was shown to be embedded in such a layer, leading to chaotic diffusion of eccentricity over gigayear timescales \cite{Tiscareno2009}. This explains the wide eccentricity distribution of 2:3 resonant KBOs, many of which show evidence of slow chaotic transport in and out of resonance. Similar structures appear near the 4:7, 5:8, and 7:12 resonances, forming a connected network of chaotic zones that strongly influences the long-term distribution of classical and scattering objects.


Chaotic transport is especially important for the production and evolution of the scattering population and Centaurs. Objects undergoing slow chaotic diffusion in eccentricity may eventually reach perihelion distances low enough for strong scattering encounters with Neptune. Once in the scattering regime, semimajor axis evolution becomes dominated by repeated planetary encounters, but the timescale for delivering KBOs to this region is governed by resonance-overlap-induced diffusion. Long-term integrations show that many scattering objects spend significant time in temporary resonances or chaotic layers before transitioning into short-period Centaur states \cite{Duncan1995,Volk2008}.

In addition to MMR overlap, overlap between secular resonances also produces chaotic behavior. In the outer classical belt, the proximity of the $\nu_8$ secular resonance to several high-order MMRs leads to combined resonance structures wherein the particle's eccentricity is modulated by both resonant and secular forcing. The resulting chaotic regions exhibit pronounced diffusion in both perihelion distance and inclination, contributing to the broad dynamical diversity observed among hot classical KBOs.

Resonance overlap also provides a natural mechanism for the formation of detached objects—those with perihelia too large to be explained by scattering alone. Objects temporarily trapped in high-order resonances may experience Kozai-like oscillations that lift  their perihelia to $q>40$ AU. Once they escape through chaotic layers, they can remain detached for gigayears \cite{Gomes2008}. This mechanism is especially efficient in resonances beyond 50~AU (e.g., 3:1, 7:2, 9:2), where resonance overlap is particularly widespread.

\begin{figure}[ht]
      \centering
      \includegraphics[width=0.8\textwidth]{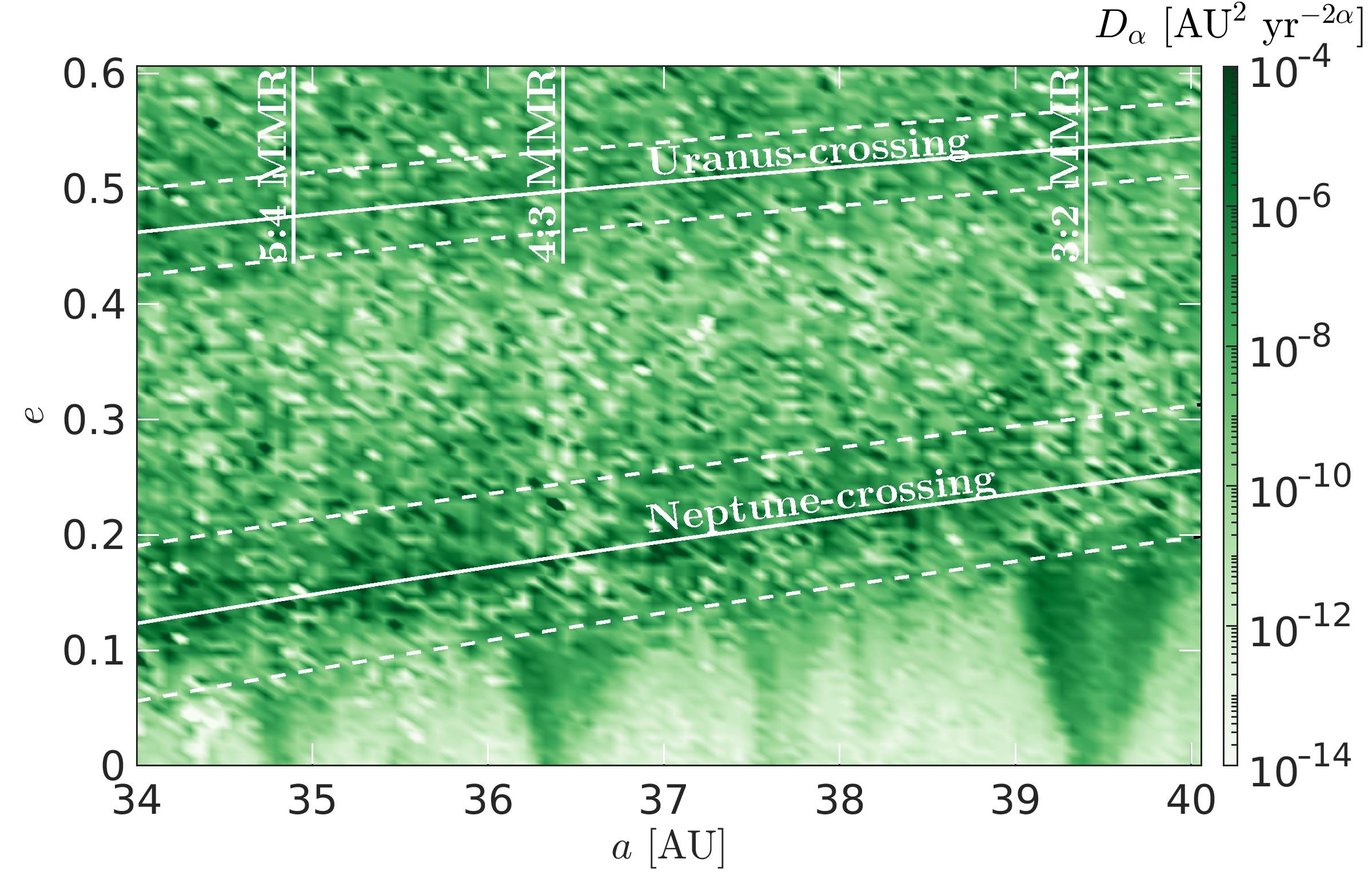}
      \caption{Depicting 
        the extended diffusion coefficients $D_\alpha$ in the $(a,e)$ plane according to mean square displacemant $\text{MSD}\,\mathbf{x}(t) = 2dD_{\alpha}t^{\alpha}$ \cite{Kovari2023},  
        where $d$ is the dimensionality of state space vector $\mathbf{x},$ $D_\alpha$ is the generalized diffusion coefficient, and the diffusion exponent $\alpha$ determines whether the process is of normal diffusion ($\alpha = 1$) or whether it is categorized as subdiffusive ($0 \leq \alpha < 1$---
      slow diffusion) or as superdiffusive ($1 < \alpha$---fast diffusion). The diffusion coefficient landscape encodes the mean-motion resonances and planet-crossing regions as the stability indicators suggest. More details in Section~\ref{sec:anom_diff}.}
      \label{fig:kovari2023_D}
\end{figure}

A comprehensive MSD-based (Mean Square Displacement) survey of chaotic diffusion in the 34-40 AU trans-Neptunian region is presented in Kővári et al. \cite{Kovari2023}, mapping the generalized diffusion coefficient $D_\alpha$ across a 100$\times$100 grid of semimajor axis and eccentricity initial conditions for 200000 test particles integrated over $2\times 10^5$ years. Figure~\ref{fig:kovari2023_D} displays the two-dimensional heat map of diffusion coefficient $D_\alpha.$ The most prominent features are the characteristic V-shaped structures of the first-order mean-motion resonances (5:4, 4:3, and 3:2 with Neptune), which attain the highest diffusion coefficients in the entire surveyed domain, reaching values of order $10^{-4}-10^{-6}$ (AU$^2$/yr)$^2$yr$^{-\alpha}$. These results (Figure~\ref{fig:kovari2023_D}) establish a quantitative framework linking local phase space structure to global instability in the inner trans-Neptunian region.

Another key result of resonance overlap is its influence on planetary migration. During Neptune's outward migration, resonances swept across the primordial disk, capturing, transporting, and releasing objects. Overlapping resonance structures enhanced transport efficiency by allowing objects to slip between resonances or escape from resonant capture. This effect is crucial for reproducing the observed distribution of resonant and non-resonant objects in population synthesis models of the Nice model \cite{Gomes2003,Levison2008}.

Frequency map analysis and chaos indicators such as Lyapunov time, MEGNO, and SALI/GALI (discussed in Section~\ref{sec:class_chaos_det}) reveal that resonance overlap dominates the chaotic structure of the Kuiper Belt. Short Lyapunov times (the inverse of Lyapunov exponents (Section~\ref{sec:class_chaos_det})) of order
\begin{equation*}
  T_{\mathrm{L}}\lesssim 10^{4}-10^{5}\text{yr}
\end{equation*}
are found in regions where MMR clusters intersect, particularly near 
$a\approx$ 42--
44 AU and $a\approx 48$ AU. These chaotic zones align with observed gaps in the distribution of non-resonant~\mbox{KBOs}.

Resonance overlap provides a comprehensive dynamical framework for understanding chaotic regions, long-term diffusion, and large-scale transport processes in the Kuiper Belt. It explains the instability of certain classical-belt regions, the origin of much of the scattering population, the dynamical structure of major resonances, the formation of detached or high-perihelion objects, and the sensitivity of the Kuiper Belt structure to Neptune's migration history. It remains one of the most powerful tools for interpreting the intricate dynamical architecture of the outer Solar System.


\section{Chaos Indicators}
\label{sec:indicators}

As we have seen above, the long-term evolution of Kuiper Belt objects (KBOs) is governed by a subtle interplay between regular and chaotic dynamics, arising from resonance overlap, secular perturbations, and close encounters with Neptune. Quantifying the degree of chaos in this region is essential for assessing orbital stability, transport pathways, and population lifetimes. Over the past decades, several complementary chaos indicators have been developed and applied extensively to trans-Neptunian dynamics.

\subsection{Classical Chaos Detection Methods}

\subsubsection{Lyapunov Exponents}
\label{sec:class_chaos_det}

The classical measure of chaos is the maximal Lyapunov exponent, defined as
\[
\lambda = \lim_{t \to \infty} \frac{1}{t} \ln \frac{\|\delta \mathbf{x}(t)\|}{\|\delta \mathbf{x}(0)\|},
\]
where $\delta \mathbf{x}(t)$ denotes the deviation between two initially nearby trajectories in phase space~\mbox{\cite{Ott2002,Benettin1980}}. Its inverse, the Lyapunov time $T_\lambda = \lambda^{-1}$, provides an estimate of the timescale over which orbital predictability is lost. In the Kuiper Belt, cold classical objects typically exhibit $T_\lambda \gtrsim 10^7$ yr, whereas scattering objects and Centaurs often have \mbox{$T_\lambda \sim 10^3$--$10^5$ yr}. Although physically intuitive, Lyapunov exponents converge slowly and require long numerical integrations, limiting their practical use in large parameter surveys~\mbox{\cite{Skokos2010}.}

\subsubsection{Mean Exponential Growth Factor of Nearby Orbits (MEGNO)}

MEGNO was introduced to accelerate chaos detection while retaining a clear dynamical interpretation \cite{Cincotta2000,Cincotta2003}. It is defined as
\[
Y(t) = \frac{2}{t} \int_0^t \frac{\|\dot{\delta \mathbf{x}}(t')\|}{\|\delta \mathbf{x}(t')\|} t' \, dt',
\]
with asymptotic behavior
\[
\langle Y \rangle \to 2 \quad \text{for regular motion}, \qquad
\langle Y \rangle \propto \lambda t \quad \text{for chaotic motion}.
\]

MEGNO converges significantly faster than Lyapunov exponents and has become a standard tool for constructing high-resolution stability maps of the Kuiper Belt \cite{Compere2013}. It efficiently delineates resonant islands, chaotic layers, and stability boundaries associated with resonance overlap.

\subsubsection{SALI and Generalized Alignment Indices}

The Smaller Alignment Index (SALI) \cite{Skokos2001} and the Generalized Alignment Index (GALI)~\cite{Skokos2007}  are numerical methods for distinguishing chaotic from regular (quasi-periodic) motion in dynamical systems. They are particularly popular for studying high-dimensional Hamiltonian systems and symplectic maps because they often identify chaos much faster than traditional Lyapunov exponents. Both methods are monitoring the alignment of multiple deviation vectors. SALI is defined as
\[
\mathrm{SALI}(t) = \min \left( \|\hat{\mathbf{w}}_1(t) + \hat{\mathbf{w}}_2(t)\|,\,
\|\hat{\mathbf{w}}_1(t) - \hat{\mathbf{w}}_2(t)\| \right),
\]
where $\hat{\mathbf{w}}_i$ are the normalized deviation vectors. For chaotic orbits, SALI decays exponentially toward zero, while for regular motion, it fluctuates around a nonzero value. GALI extends this concept to higher-dimensional phase space by considering the volume spanned by multiple deviation vectors. These indices are particularly effective in distinguishing weak chaos from regular motion and converge rapidly, making them well-suited for detailed analysis of resonant regions in the Kuiper Belt.

\subsubsection{Frequency Map Analysis}

Frequency map analysis provides a complementary, quasi-integrable perspective on orbital stability by decomposing motion into fundamental frequencies. For a regular orbit, the dominant frequencies remain constant in time, whereas chaotic orbits exhibit slow diffusion in frequency space \cite{Laskar1990,Laskar1993}. By tracking variations such as
\[
\Delta \nu = |\nu_1 - \nu_2|,
\]
over successive time intervals, frequency analysis quantifies diffusion rates and identifies resonant structures. In Kuiper Belt studies, frequency maps have proven especially powerful for identifying resonance overlap, secular resonances, and the boundaries of long-term~\mbox{stability}.


\subsubsection{Maximum Eccentricity Method}

A conceptually simple yet dynamically informative stability indicator is the 
maximum eccentricity method 
(MEM) \cite{Dvorak2005}, which characterizes orbital stability 
through the time evolution of the osculating eccentricity $e(t)$. 

For a given initial condition $\mathbf{x}_0$, one numerically integrates the 
equations of motion over a prescribed time interval $[0,T]$ and records

\begin{equation}
e_{\max}(\mathbf{x}_0) = \max_{0 \le t \le T} e(t).
\end{equation}

\begin{figure}[ht]
      \centering
      \includegraphics[width=0.8\textwidth]{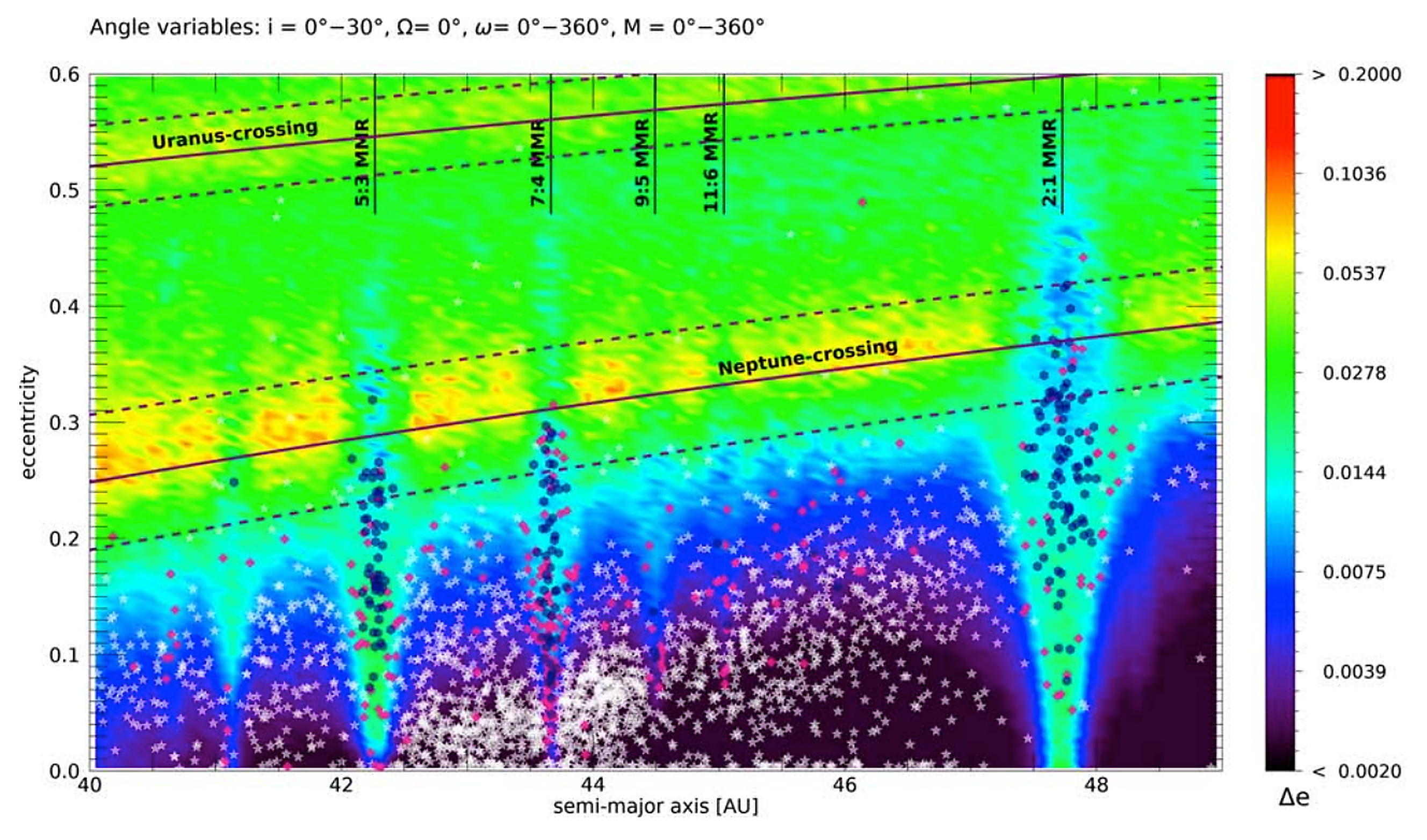}\\
      \caption{Stability 
      map of the TNO region based on the maximum eccentricity method. Non-resonant (white), short-term (pink), and long-term (blue) resonant real TNOs are marked to show the dynamical importance of mean-motion resonances. The Neptune- and Uranus-crossing orbits are also shown (solid lines). The three Hill radii distances to the giant planets are also marked (dashed lines). Source of figure: \cite{Forgacs2023}.}
      \label{fig:fde2024_de}
\end{figure}

The quantity $e_{\max}$ serves as a proxy for dynamical stability. 
If the orbit remains confined to a quasi-periodic torus or a stable 
mean-motion resonance, the eccentricity typically exhibits bounded 
oscillations.
Conversely, in chaotic regions where resonance overlap occurs, eccentricity 
may undergo stochastic diffusion,

As demonstrated in Figure~\ref{fig:fde2024_de} in the Kuiper Belt context, the maximum eccentricity method is especially useful for mapping stability boundaries near Neptune's mean-motion resonances. Objects trapped in stable resonant islands maintain moderate eccentricities, while trajectories in chaotic layers may experience secular pumping that drives $e_{\max}$ toward unity. The method therefore provides a fast diagnostic for identifying unstable initial conditions without requiring the computation of variational equations or chaos indicators such as Lyapunov exponents.



Together, these chaos indicators provide a robust and complementary toolkit for characterizing the complex dynamical landscape of the Kuiper Belt. Lyapunov exponents offer a fundamental measure of chaos, MEGNO enables rapid global mapping, SALI/GALI capture fine-scale instability, and frequency analysis reveals the underlying resonant architecture. Their combined application is essential for understanding stability, transport, and population evolution in the trans-Neptunian region.

\subsection{Additional Contemporary Stability Indicators}
\label{sec:new_chaos_det}

In recent years, several alternative and complementary stability indicators have been developed to address specific limitations of classical chaos diagnostics, particularly in the context of resonant dynamics, weak chaos, and long-term transport. Among these, the FAIR method, entropy-based indicators, and anomalous diffusion diagnostics have proven especially valuable for the analysis of Kuiper Belt dynamics.
\vspace{-6pt}

\subsubsection{Lagrangian Descriptors}

Lagrangian Descriptors (LDs) \cite{Daquin2022} describe a geometric approach to identifying phase-space structures by integrating a scalar quantity along trajectories. A commonly used LD is
\begin{equation*}
      \mathrm{LD}(\mathbf{x_0})=\int_{-T}^{T}||\dot{\mathbf{x}}(t;\mathbf{x_0})||\mathrm{d}t,
\end{equation*}
where $\mathbf{x_0}$ is the initial condition. Sharp gradients in LD maps reveal invariant manifolds, separatrices, and transport channels \cite{Caliman2025}.

Recent work by Daquin et al. \cite{Daquin2023} demonstrated that LDs are particularly effective for visualizing transport pathways in deterministic chaos and can be applied to the Kuiper Belt objects. LD maps clearly delineate:
(i) stable resonant islands; (ii) chaotic layers around resonances; (iii) pathways connecting resonant regions to the scattering disk.

\begin{figure}[ht]
      \centering
      \includegraphics[width=0.8\textwidth]{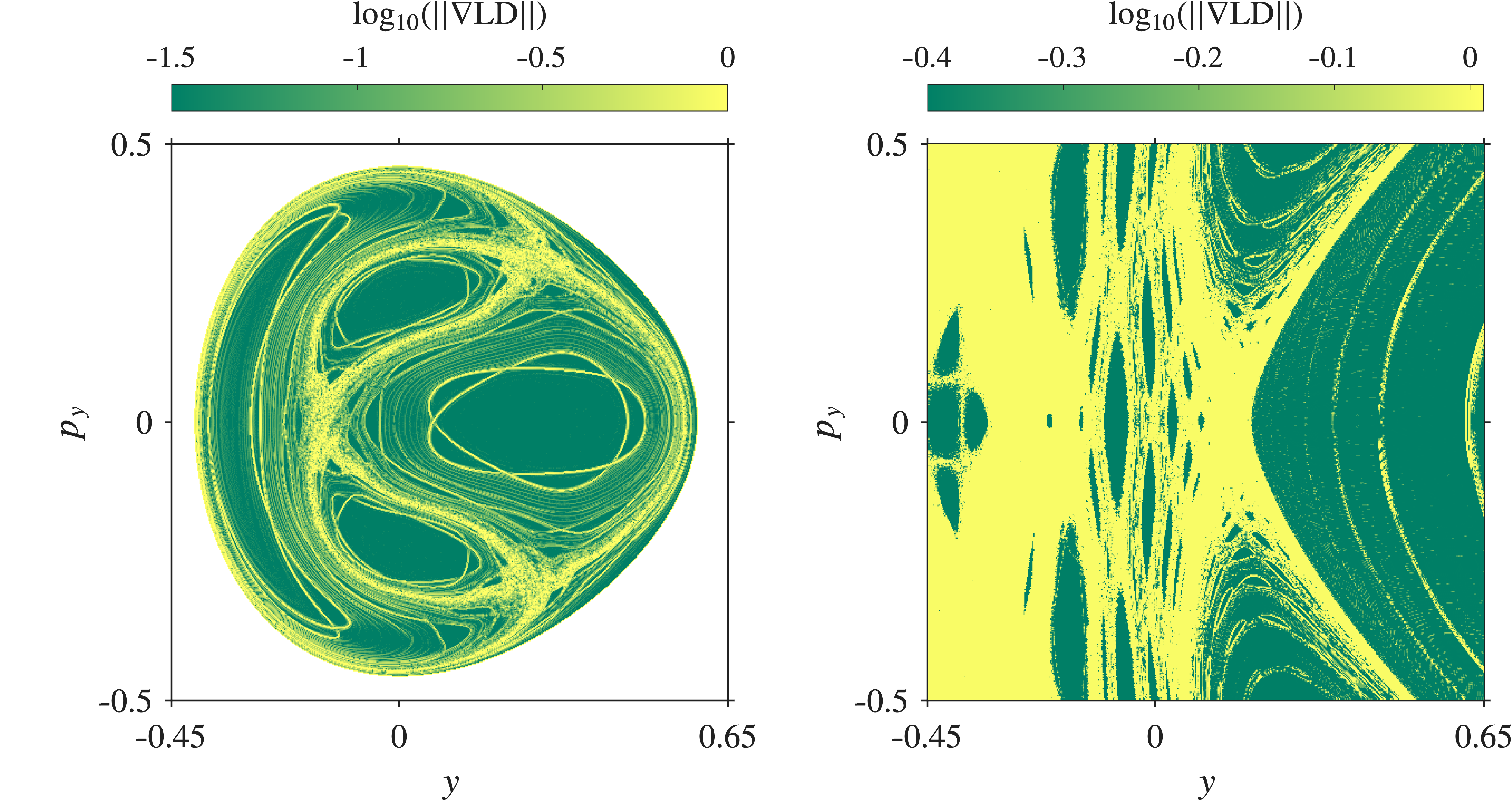}
      \caption{Dynamical 
 map of Poincaré section ($p,p_y$) in the Hénon-Heiles system with the Lagrangian Descriptor indicator. LD represents the accurate texture of the dynamics, including invariant curves, chaotic bands, and sticky regions. Source of figure: \cite{Daquin2022}.}
      \label{fig:jerome2022} 
\end{figure}

Unlike traditional chaos indicators, LDs emphasize the global geometry of phase space, Figure~\ref{fig:jerome2022},  making them especially useful for understanding resonance-driven transport and the origin of scattering and detached populations.

\subsubsection{FAIR: Fast Identification of Mean-Motion Resonances}

The Fast identification of mean-motion resonances (FAIR) method, introduced by Forgács-Dajka and Sándor \cite{Forgacs2018}, is designed to efficiently detect and characterize mean-motion resonances without requiring long-term integrations or computation of variational equations. FAIR exploits the fundamental property of resonant motion: the commensurability between orbital frequencies.

\begin{figure}[ht]
      \centering
      \includegraphics[width=0.8\textwidth]{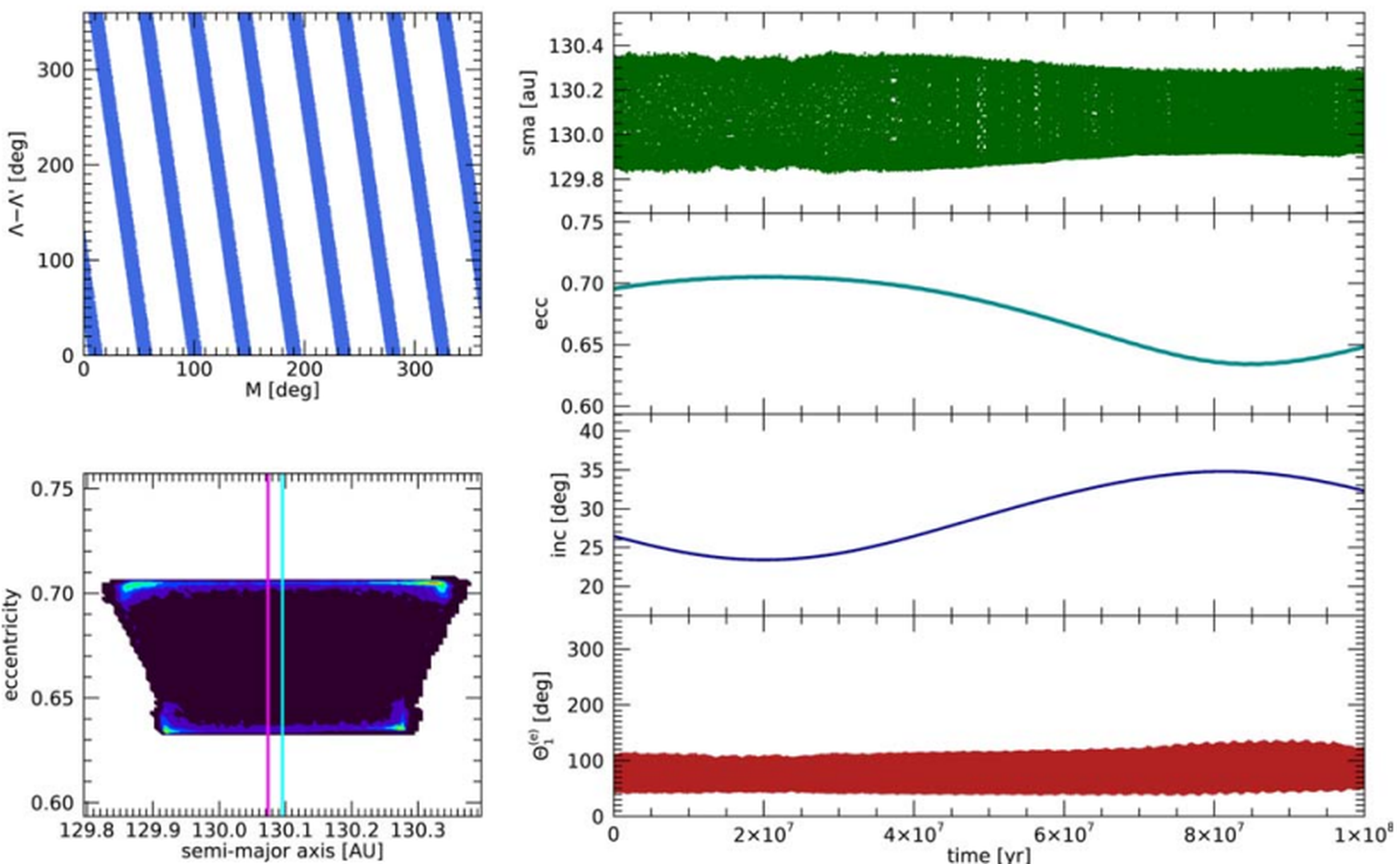}
      \caption{The 
 asteroid $2007\,\mathrm{TC}_{434}$ captured in 9:1 MMR with Neptune  \cite{Forgacs2022}. The upper left panel shows the $\Lambda - \Lambda'$ vs. $M,$ the difference of mean longitudes and mean anomaly, respectively, plane used to determine the type, order, and degree of the resonance by applying the FAIR method. The lower left panel reveals the region covered by the asteroid in the $(a, e)$ plane during the whole length of the numerical integration. The right panels display, respectively, the time evolution of the orbital elements $a,$ $e,$ and $i,$ and the critical argument of the eccentricity-type MMR.}
      \label{fig:fde2024_fair}
\end{figure}

For a test particle with mean longitude $\lambda$ and a perturbing planet with mean longitude $\lambda_p$, the method examines combinations of the form
\[
\phi_{p:q} = p \lambda - q \lambda_p - (p-q)\varpi,
\]
and evaluates the temporal behavior of these angles over relatively short integrations. Rather than tracking libration directly, FAIR analyzes the time series of $\dot{\lambda}$ and constructs frequency ratios
\[
\frac{n}{n_p} \approx \frac{p}{q},
\]
where $n$ and $n_p$ are the mean motions of the particle and planet \cite{Forgacs2022}, respectively (see Figure~\ref{fig:fde2024_fair}).

The key advantage of FAIR lies in its ability to detect resonant behavior through clustering in frequency-ratio space. Thus, FAIR rapidly maps resonant structures across large regions of phase space. This makes it particularly well-suited for surveys of the Kuiper Belt, where numerous high-order and weak resonances coexist.

FAIR is computationally inexpensive and robust against short-term perturbations, but it does not directly quantify chaoticity. As such, it is most effective when used in combination with chaos indicators such as MEGNO or frequency diffusion analysis, providing a rapid first-pass classification of resonant versus non-resonant motion.

\subsubsection{Entropy-Based Chaos Indicators}

Entropy-based methods provide a statistical measure of orbital complexity by quantifying the information content or disorder in time series derived from orbital elements \cite{Cincotta2021b}. These approaches are rooted in information theory and are particularly sensitive to weak chaos and long-term diffusion \cite{Frigg2004}.

One commonly used measure is the Shannon entropy,
\[
S = -\sum_{i} p_i \ln p_i,
\]
where $p_i$ represents the probability of the system occupying a given region of phase space or frequency bin. In practical applications, the phase space is discretized using orbital elements or frequency components extracted from time series data.

\begin{figure}[ht]
      \centering
      \includegraphics[width=0.8\textwidth]{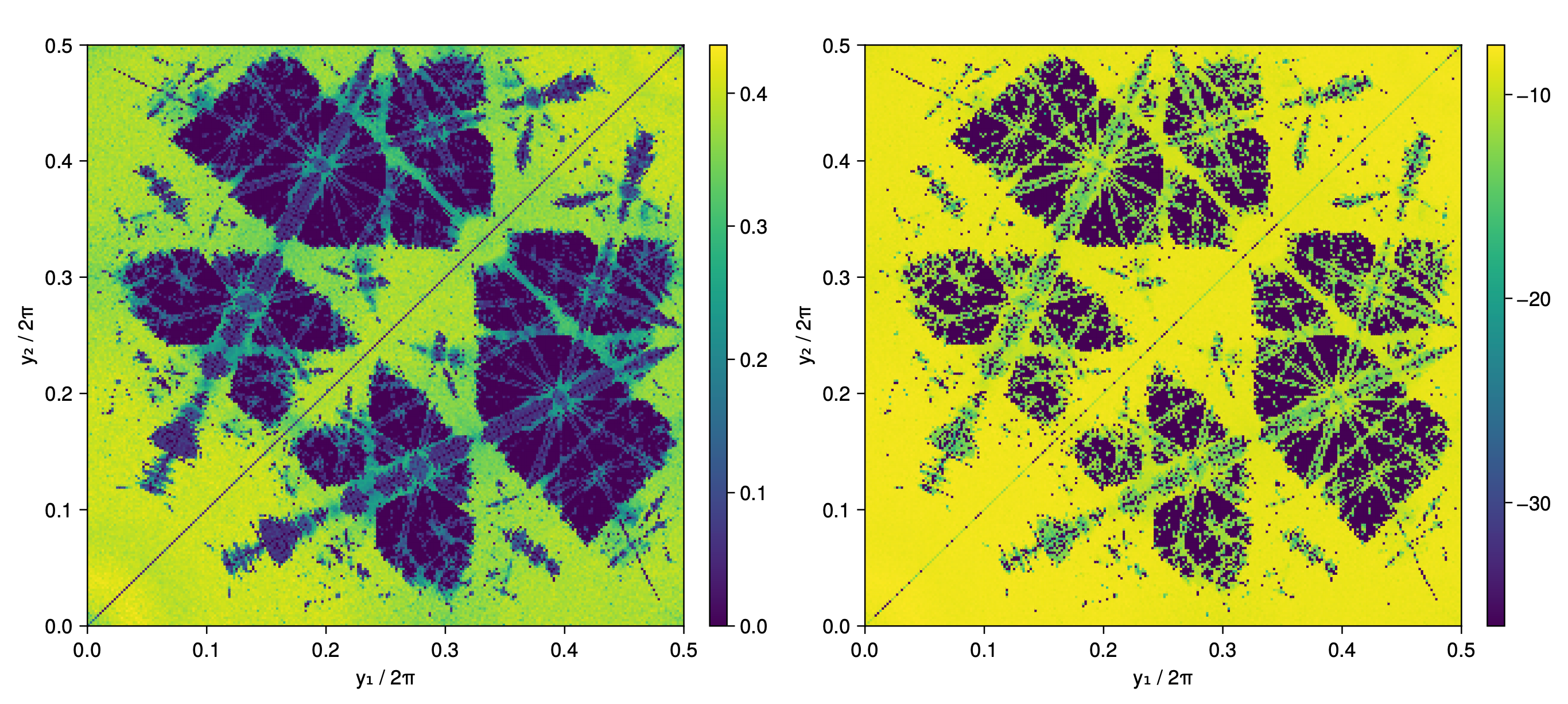}
      \caption{Phase 
 space portrait of the 4D Hamiltonian resonance web (see Equation~(1) in \cite{Giordano2018}) for parameters $\epsilon=0.25,$ $\gamma=0.1,$ $\mu=0.5.$ Contour plot for $S$ (\textbf{left}) and $S'\sim \mathrm{d}S/\mathrm{d}t$ (\textbf{right}, in logarithmic scale) for a grid of 500 $\times$ 500 initial conditions after $t = 5 \times 10^5.$ The figure is reproduced from \cite{Giordano2018}.}
      \label{fig:giordano2018} 
\end{figure}

For regular or quasi-periodic motion, the system explores a limited region of phase space, resulting in low entropy values; see Figure~\ref{fig:giordano2018}. In contrast, chaotic trajectories exhibit broader phase-space exploration and higher entropy \cite{Lesne2014}. Cincotta and collaborators demonstrated that entropy growth rates provide a reliable proxy for dynamical instability, often converging faster than Lyapunov exponents \cite{Cincotta2021a,Cincotta2023}.

\begin{figure}[ht]
      \centering
      \includegraphics[width=0.8\textwidth]{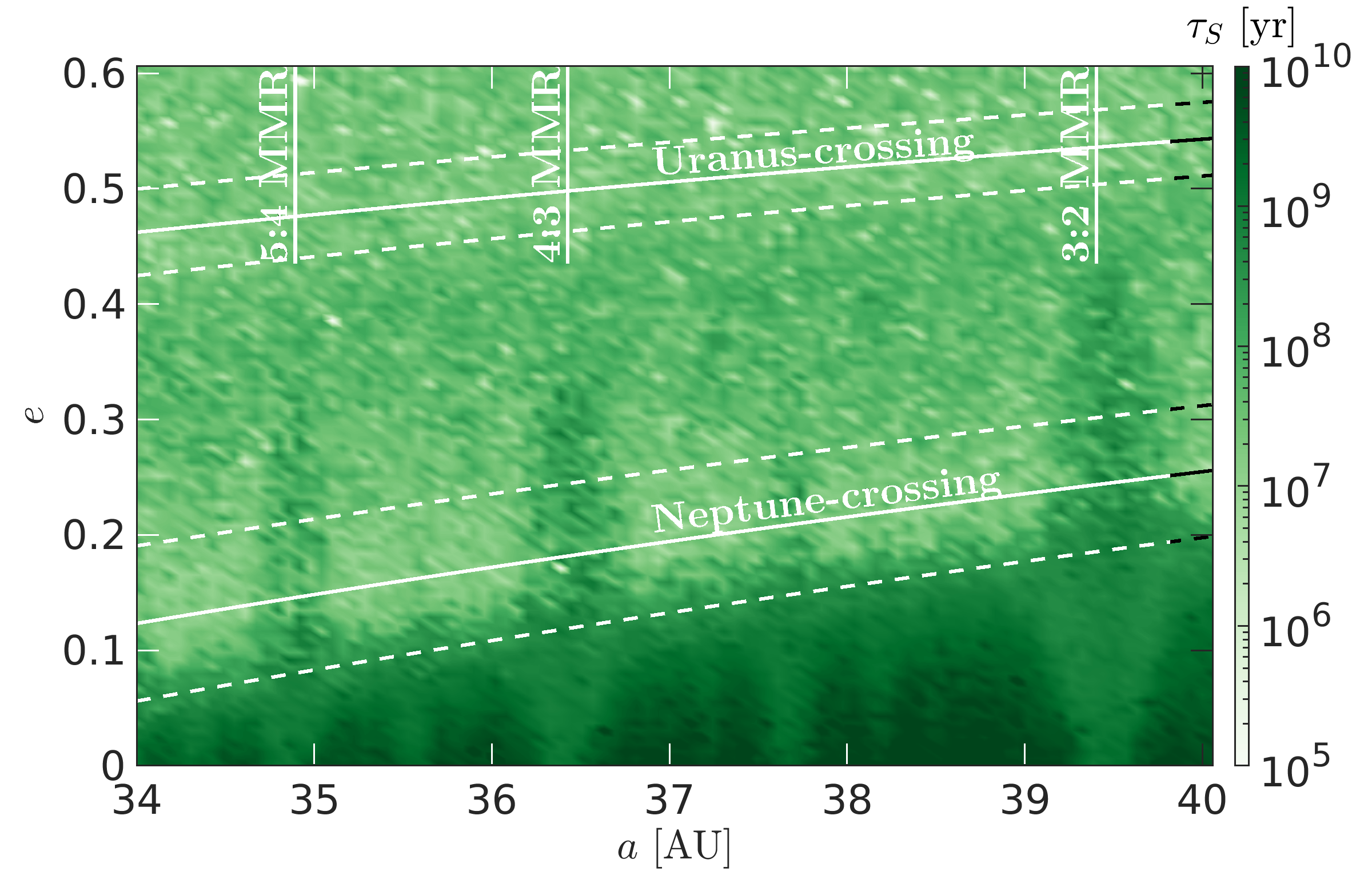}
      \caption{Heat map of the stability times computed from the Shannon entropy based on $\mathbf{x} = (L, G, H)$ in the 34-40 AU region of the trans-Neptunian space (between the eccentricities $0 \leq e \leq 0.6$). Here, $L,\,G,\,H$ denote the Delaunay actions (of dimension AU$^2$ yr$^{-1}$). The $(a, e)$ pairs drawing solid lines result in Neptune- or Uranus-crossing orbits. Dashed lines: three Hill radius distances from the two giant planets. Vertical lines on the top of the figure indicate the location of MMRs. Source: \cite{Kovari2022}.}
      \label{fig:kovari2023_tauSh}
\end{figure}

The Shannon entropy formalism has recently been extended to
multiplanet systems by Kővári et al. \cite{Kovari2022},
who applied it to the planar non-restricted four-body problem
in the context of resonant exoplanetary dynamics.
Projecting the motion onto an action space (e.g., $a,e$),
the entropy of a trajectory $S(t)$ is computed as
\begin{equation}
S(t) = \ln N - \frac{1}{N}\sum_{k=1}^{r} n_k \ln n_k,
\end{equation}
where $N$ is the number of intersections with a defined section
and $n_k$ the number of visits to the $k$-th cell of a partition.
Regular trajectories rapidly saturate at low normalized entropy,
whereas chaotic trajectories exhibit sustained entropy growth.
Importantly, the time derivative $dS/dt$ was related to a diffusion
coefficient in action space \cite{Giordano2018},
\[
D_S \propto \frac{dS}{dt},
\]
allowing an estimate of characteristic stability times
$\tau_{\mathrm{s}} \sim 1/D_S$ (Figure~\ref{fig:kovari2023_tauSh}).
This provides a quantitative link between entropy growth and
chaotic transport.

Kovács et al. \cite{Kovacs2022} further generalized the approach by introducing
the Rényi entropy of order $\alpha$,

\begin{equation}
S_\alpha = \frac{1}{1-\alpha}
\ln \left( \sum_{k=1}^{r} p_k^\alpha \right),
\end{equation}
where $p_k = n_k/N$.
The Shannon entropy is recovered in the limit $\alpha \to 1$.
By varying $\alpha$, the method becomes sensitive to different
features of phase-space exploration, improving the discrimination
between weak chaos, sticky motion, and strongly diffusive
trajectories.

Entropy-based diagnostics therefore complement Lyapunov-type
indicators and are particularly suitable for resonance-dominated
systems such as trans-Neptunian populations, where slow chaotic
diffusion governs long-term stability.

\subsubsection{Anomalous Diffusion and Transport Timescales}
\label{sec:anom_diff}

Traditional diffusion models assume a linear growth of variance with time,
\[
\langle (\Delta a)^2 \rangle \propto t,
\]
corresponding to normal (Brownian) diffusion. However, chaotic transport in the asteroid belt(s) often deviates from this behavior due to resonance sticking and hierarchical phase-space structure \cite{Cordeiro2006,Tsiganis2007}.

Kovári et al. \cite{Kovari2023} introduced a framework for characterizing anomalous diffusion, in which the variance follows a power-law relations,
\[
\langle (\Delta a)^2 \rangle \propto t^{\alpha},
\]
with $\alpha \neq 1$. Values $\alpha < 1$ indicate subdiffusion associated with long trapping times in resonances, while $\alpha > 1$ corresponds to superdiffusion driven by intermittent close encounters or resonance overlap.

From this formalism, an anomalous diffusion timescale $T_D$ related to $D_{\alpha}$ in Figure~\ref{fig:kovari2023_D} can be defined as the characteristic time required for an orbit to undergo macroscopic transport across a specified region of phase space. These timescales provide a physically meaningful measure of long-term stability, particularly for objects whose instantaneous chaoticity does not directly translate into rapid orbital evolution \cite{Cincotta2022,Shevchenko2020}.

\subsubsection{Phase Space Divergence Based on Poincar\'e Recurrences} 

\begin{figure}[ht]
      \centering
      \includegraphics[width=0.85\textwidth]{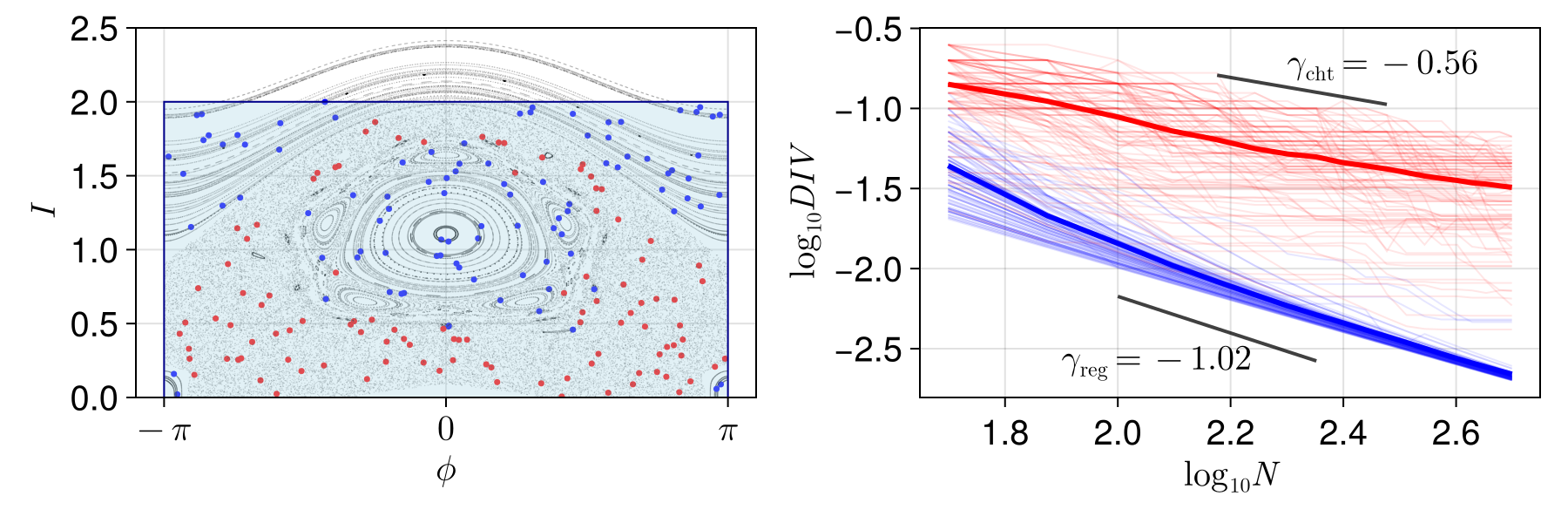}
      \caption{(\textbf{Left}) Phase 
      portrait for the resonance overlap Hamiltonian \cite{Daquin2026} obtained with its associated Poincar\'e map. (\textbf{Right}) The ensemble averages of the chaos indicator DIV follow distinct power laws on the regular (blue) and chaotic (red) components. The initial conditions of the ensemble members are marked in the left panel (blue dots - regular, red dots - chaotic motion). Source of the figure: \cite{Daquin2026}.}
      \label{fig:jerome2025}
\end{figure}

The recent study introduces a chaos detection method based on the divergence of recurrence plots, extending classical recurrence quantification analysis (RQA) \cite{Zbilut1992,Webber1994} by explicitly measuring the exponential separation of nearby trajectories in reconstructed phase space; see more details in Appendix 
\ref{sec:appA}. Rather than relying on tangent-space dynamics or variational equations, the method extracts a divergence rate from the statistics of diagonal line structures in recurrence plots, which encode the temporal persistence of phase-space proximity. It has been demonstrated (Figure~\ref{fig:jerome2025}) that the resulting divergence indicator correlates closely with the Fast Lyapunov Indicator \cite{Froeschle1997} for both continuous-time and discrete-time dynamical systems, supporting the reported \cite{Marwan2007} correlation between the divergence of recurrences and positive Lyapunov exponents.
Importantly, the method remains effective in regimes of weak chaos where traditional indicators suffer from slow convergence, and it does not require explicit knowledge of the governing equations \cite{Takens1981,Small2004,Hirata2006}. These properties make recurrence plot divergence particularly suitable for applications in celestial mechanics, including the analysis of resonant Kuiper Belt objects and observationally constrained orbital data, where only limited time series are available, and long integrations are impractical.

\subsection{Applications to Kuiper Belt Populations}

The combined use of stability indicators has led to several key insights into Kuiper Belt dynamics:

\begin{enumerate}
      \item Cold classical KBOs are confirmed to occupy the most stable regions of phase space, with minimal chaotic diffusion.
      \item Resonant populations exhibit a spectrum of stability, from deeply stable librators to weakly chaotic, long-lived objects.
      \item Scattering objects are dominated by strong chaos, with short Lyapunov times and rapid diffusion.
      \item Detached objects often lie in weakly chaotic regions shaped by resonance overlap and secular effects.
      \item Resonance sticking, identified through chaos indicators, is a major contributor to the long-term residence of objects near Neptune.
\end{enumerate}
Stability indicators are also crucial for validating numerical models of Neptune’s migration. Successful models must reproduce not only the observed orbital distributions but also the inferred stability structure revealed by chaos diagnostics.

In the Kuiper Belt, anomalous diffusion analysis has revealed that many resonant and near-resonant objects experience extended periods of quasi-stability punctuated by rapid transitions, a behavior that cannot be captured by single-value chaos indicators alone.

\medskip
Together, FAIR, entropy-based indicators, and anomalous diffusion diagnostics enrich the toolbox of modern celestial mechanics. FAIR enables rapid identification of resonant structures, entropy-based methods quantify weak chaos and complexity, and anomalous diffusion analysis connects chaotic dynamics with long-term transport timescales. Their combined application provides a more complete and nuanced characterization of stability and evolution in the Kuiper Belt than any single indicator in isolation.

\subsection{Outlook}

As observational surveys expand the known Kuiper Belt population by orders of magnitude, stability indicators will become increasingly important for automated classification and interpretation. Coupled with machine learning approaches (see Section~\ref{sec:ml}), chaos indicators will enable rapid assessment of orbital stability, identification of resonant membership, and reconstruction of the dynamical history of the outer Solar System.

Stability indicators provide the quantitative backbone of modern Kuiper Belt dynamics, transforming numerical integrations into physical insight and linking observed orbital structure to the fundamental mechanisms of resonance, chaos, and planetary migration.


\section{Machine Learning-Based Gravity and Dynamical Models}
\label{sec:ml}

The long-term dynamical evolution of Kuiper Belt objects is traditionally investigated through direct numerical integration of the $N$-body problem. While this approach is robust, it becomes computationally expensive when large ensembles or extended timescales are required. Recent studies have demonstrated that machine learning (ML) surrogate models can serve as efficient approximations to gravitational dynamics when trained on high-precision numerical solutions.

\subsection{Machine Learning Surrogate Gravity Models}
\label{sec:ml_surrogate}

\begin{figure}[ht]
      \centering
      \includegraphics[width=0.85\textwidth]{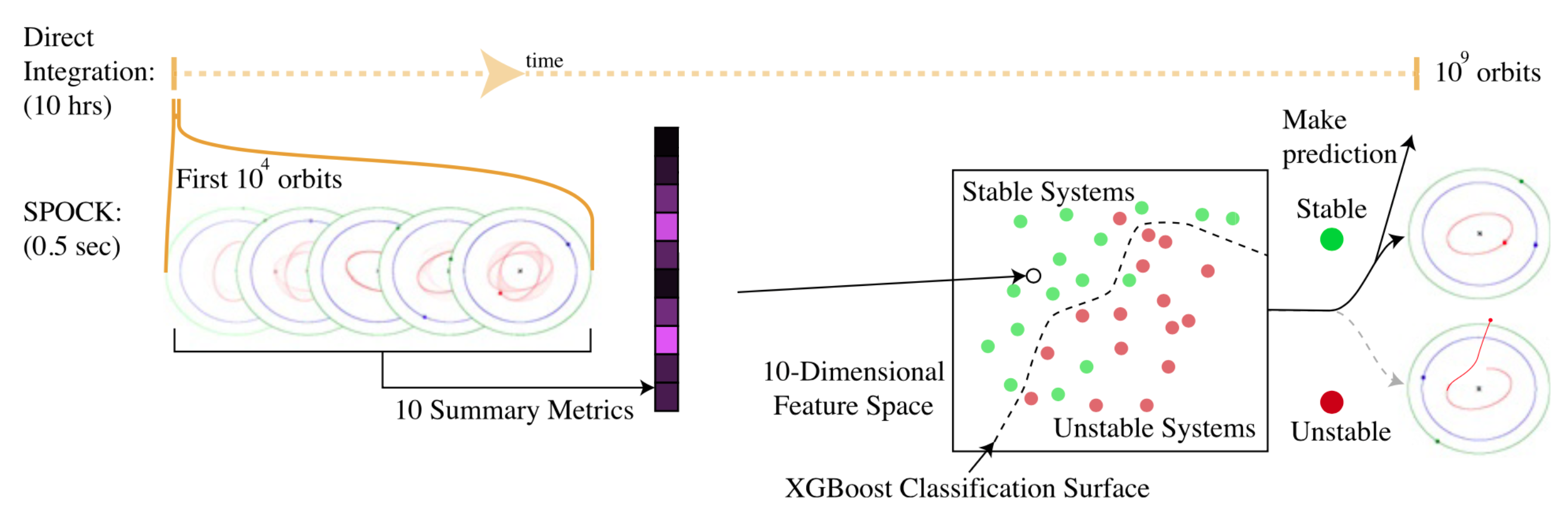}
      \caption{A 
      cartoon of the SPOCK (Stability of Planetary Orbital Configurations Klassifier) workflow. The method involves supervised learning (XGBoost) based on a 10-dimensional feature parameter space. The training data is a set of short-term integration of closed three-planet systems. The decision whether a system is stable for $10^9$ orbits is $10^5$ times faster with this algorithm than direct integration. Source of the figure: \cite{Tamayo2020}.}
      \label{fig:tamayo2021}
\end{figure}

Recently, the SPOCK (Stability of Planetary Orbital Configurations Klassifier) method, a machine learning framework \cite{Tamayo2020}, was introduced for classifying the long-term dynamical stability of compact multiplanet systems over $10^9$ orbital timescales. The central innovation lies in replacing computationally prohibitive full N-body integrations with a gradient-boosted decision tree (XGBoost; \cite{Chen2016}) trained on physically motivated summary statistics extracted from short $10^4$-orbit integrations, yielding speed-ups of up to $10^5.$ The model is trained on ~100,000 three-planet configurations sampled preferentially in and near mean-motion resonances (MMRs), motivated by analytical work demonstrating that MMR overlap is the dominant driver of rapid dynamical instabilities in closely packed systems \cite{Wisdom1980,Deck2013,Quillen2011}. During the process, a bunch of summary features is constructed from resonant orbital dynamics, including the MEGNO chaos indicator, normalized eccentricity mode variances, and MMR strength metrics, all evaluated within the short integration window (Figure~\ref{fig:tamayo2021}). Generalization tests on uniformly distributed and higher-multiplicity configurations confirm that the model captures physically meaningful dynamics rather than memorizing training-set particulars, supporting the hypothesis that MMR-driven instabilities are local and well-approximated by considering adjacent planetary trios \cite{Lithwick2012}. It has also been shown that training on resonant configurations generalizes robustly to non-resonant phase space, providing empirical confirmation that short-timescale instabilities in compact systems are predominantly MMR-driven, while secular mechanisms (e.g., \cite{Lithwick2011}) govern longer-timescale destabilization.
\vspace{-6pt}



Breen et al. \cite{Breen2020} demonstrate that a deep artificial neural network (ANN) can produce accurate solutions to the planar chaotic three-body problem over a bounded time interval, operating up to $10^8$ times faster than the arbitrary-precision Brutus integrator \cite{Boekholt2015} upon which its training data are based. The authors restrict their proof-of-concept to three equal-mass particles with zero initial velocities arranged in a symmetric planar configuration, reducing the general solution to a mapping from a three-dimensional input space (time and one initial coordinate) to particle positions. A feed-forward ANN with 10 hidden layers of 128 nodes, optimized using the ADAM algorithm \cite{Kingma2015}, was trained on ~9900 converged trajectories and validated on a held-out set of 100, achieving a mean absolute error  $\leq 0.1.$ A critical methodological finding is that training on arbitrarily precise converged solutions, rather than conventional double-precision integrations, is essential, as fixed-precision integrators can introduce systematic errors near close encounters that are subsequently learned and propagated by the network. The ANN was further shown to faithfully reproduce the hallmark sensitive dependence on initial conditions of chaotic systems, as quantified through estimated Lyapunov exponents across 4,000 trajectory pairs. Energy conservation, typically an implicit guarantee of symplectic integrators, was addressed through a post-hoc projection layer that reduces relative energy errors from $~10^{-2}$ to $~10^{-5}.$ 

These results \cite{Breen2020} highlight that, over finite time intervals, deep neural networks can serve as accurate surrogates for traditional N-body integrators, even in highly chaotic regimes. Crucially, this accuracy is achieved at a fixed and dramatically reduced computational cost, orders of magnitude faster than direct numerical integration, suggesting strong potential for large-scale simulations in celestial mechanics.

In the context of Kuiper Belt dynamics, surrogate models are particularly useful for accelerating ensemble studies of resonance capture, short-term diffusion, and local stability. Physics-informed architectures that preserve Hamiltonian structure or approximate symplecticity are especially promising, given the near-integrable nature of trans-Neptunian motion. ML surrogates should therefore be regarded as complementary tools that enhance computational efficiency while remaining grounded in classical celestial mechanics.

\subsection{Machine Learning for Chaos Detection and Stability Classification}
\label{sec:ml_chaos}

Machine learning techniques have also been applied successfully to the classification of orbital stability and chaos in Hamiltonian systems. Lemos et al.~\cite{Lemos2023} demonstrated that supervised ML classifiers trained on short integrations can distinguish regular and chaotic trajectories with accuracy comparable to classical chaos indicators such as Lyapunov exponents and MEGNO.

\begin{figure}[ht]
      \centering
      \includegraphics[width=0.85\textwidth]{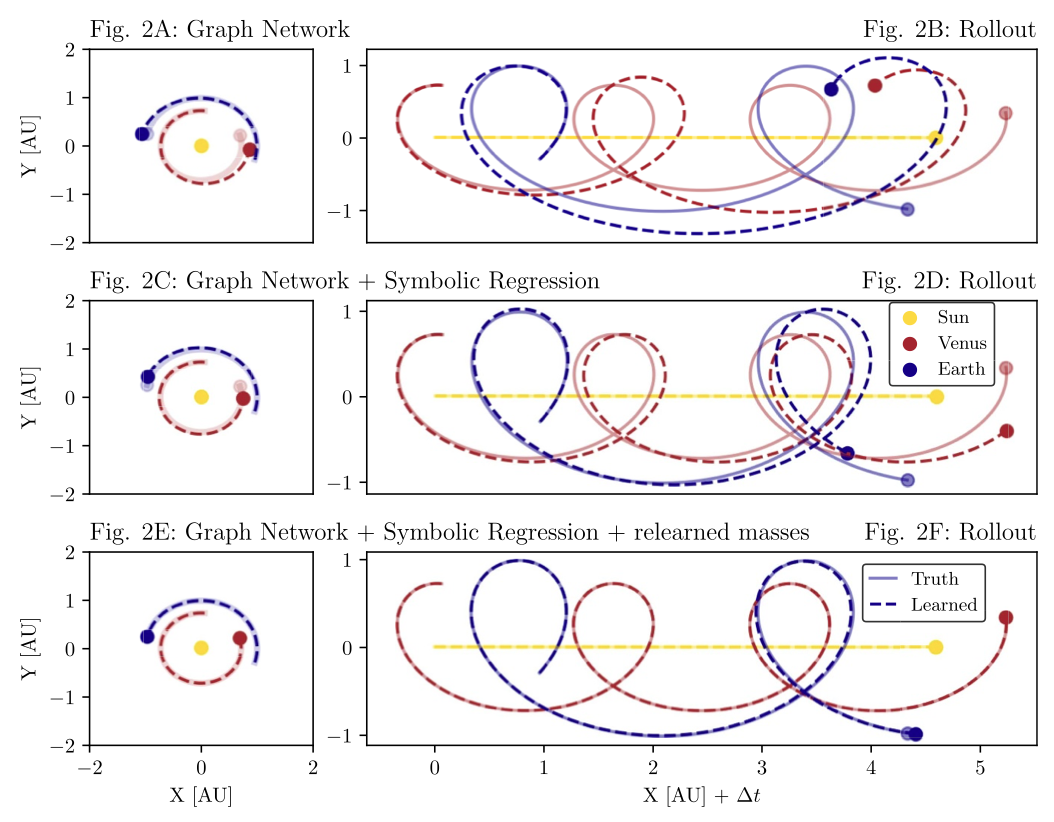}
      \caption{Graph 
      networks [GN] (nodes and edges) trained by real observational data of the Solar System objects (sun, planets, and moons). By optimizing the parameters of the GN neural network edge functions, it is possible to fit the force expression, which, in the planetary case, is Newton's law of gravitation. Having the equations of motion from the fit, refined masses can be obtained that match well the actual masses in the Solar System. Each row shows a different stage of the method. The left column depicts trajectories for six months, while the right column follows the motion until 21 months. Source of the figure: \cite{Lemos2023}.}
      \label{fig:lemos2023}
\end{figure}

Figure~\ref{fig:lemos2023} illustrates the progressive improvement in modeling Solar System dynamics achieved by the proposed two-step approach. The graph neural network (GN) alone accurately reproduces short-term orbital trajectories, but its predictions diverge over longer time horizons due to error accumulation inherent in iterative rollouts. When symbolic regression is applied to extract an explicit force law from the GN, the resulting analytical model yields noticeably more stable and accurate long-term trajectories, demonstrating superior generalization despite its lower complexity. Furthermore, re-estimating the planetary masses using the discovered law leads to an additional improvement, producing trajectories that closely match the ground truth over extended periods. The results highlight a clear progression from a purely data-driven simulator to an interpretable, physically grounded model, showing that the combination of neural networks and symbolic regression not only recovers the correct governing law but also enhances predictive accuracy and~\mbox{stability}.

\subsection{Data-Driven Reconstruction of Dynamical Stability}
\label{sec:ml_data_driven}

Machine learning also offers new possibilities for reconstructing planetary dynamics from the observed structure of the system. The imprint of Neptune's migration is encoded in resonant populations, inclination distributions, and kernel-like features in the Kuiper Belt, but the inverse problem remains highly degenerate.

Hybrid approaches that combine physics-based models with data-driven corrections have proven effective in chaotic systems. Pathak et al.~\cite{Pathak2018} introduced a hybrid forecasting framework in which an imperfect physical model is augmented by an ML component trained to correct systematic errors. This approach can be suited to migration modeling, where simplified prescriptions fail to capture stochastic effects such as resonance sticking.

\begin{figure}[ht]
      \centering
      \includegraphics[width=0.8\textwidth]{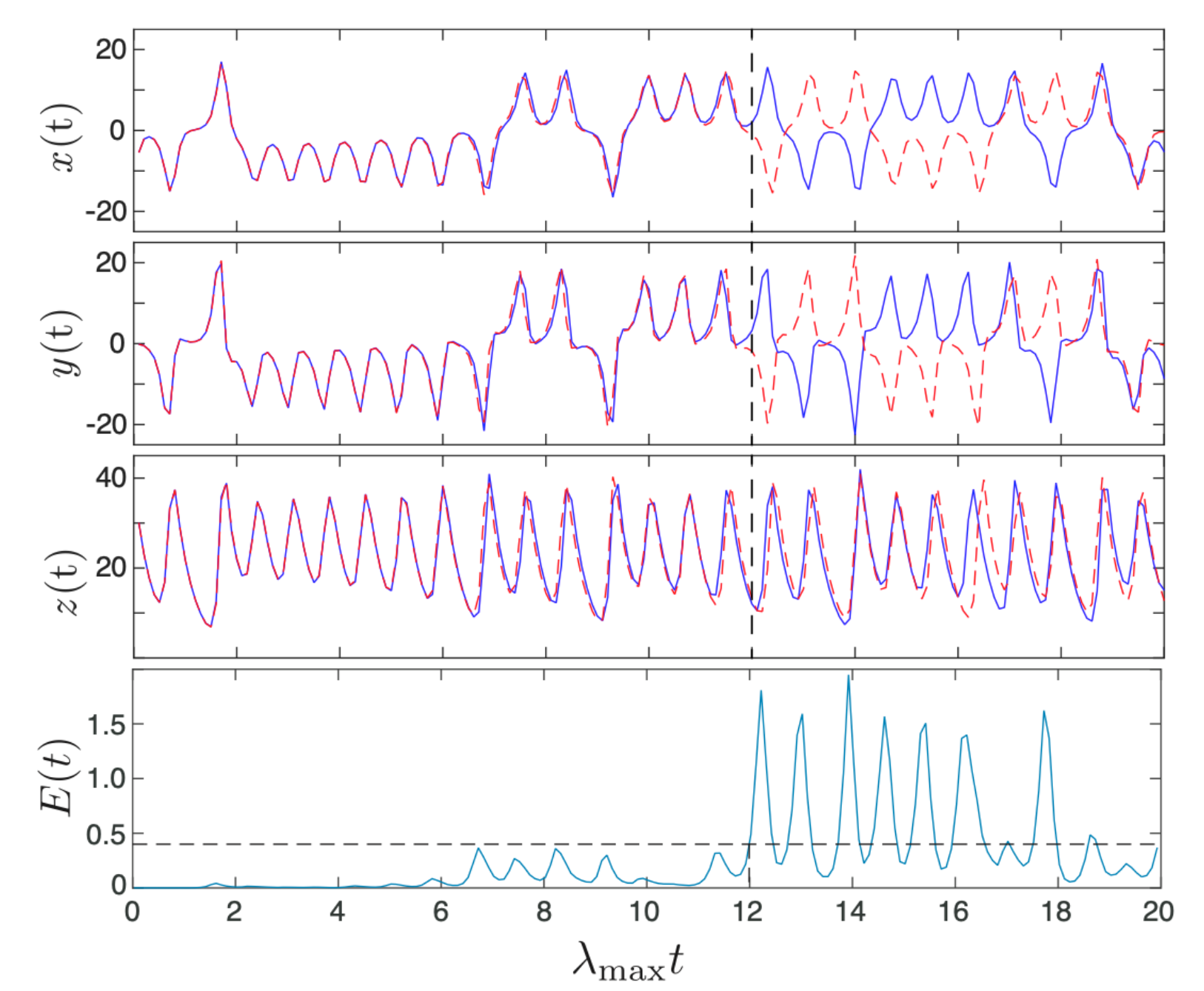}
      \caption{Upper 
 3 panels: Prediction of the Lorenz system \cite{Lorenz1963} using the reservoir computing model with the extension of the knowledge-based predictor loop. The blue line shows the ground truth of a strongly chaotic system, and the red dashed line indicates the hybrid prediction. Lower panel: The normalized prediction error remains below the pre-defined threshold (dashed line) for about 12 Lyapunov times. Source of the figure: \cite{Pathak2018}.}
      \label{fig:pathak2018}
\end{figure}

Figure~\ref{fig:pathak2018} demonstrates the predictive capability of the proposed hybrid forecasting framework for chaotic dynamics, using the Lorenz system as a test case. The results show that the hybrid model combining a reservoir computing approach \cite{Lu2018,Krishnagopal2020} with an imperfect knowledge-based model can accurately track the true system trajectory for approximately 10-12 Lyapunov times before divergence, significantly outperforming either component used in isolation. In particular, the hybrid prediction maintains low normalized error well beyond the validity horizon of both the standalone machine learning model and the inaccurate physical model, illustrating how the combination effectively compensates for deficiencies. This improvement is further quantified by the extended “valid time,” indicating that the hybrid approach substantially delays the onset of exponential error growth characteristic of chaotic systems.

These results suggest a promising application to the dynamics of trans-Neptunian small bodies, where orbital evolution is inherently chaotic and often modeled with incomplete physics or uncertain initial conditions. A hybrid forecasting framework could leverage approximate N-body models together with observational time series to extend reliable prediction horizons for such objects. In particular, it may improve the tracking of resonant or weakly interacting bodies, where small modeling errors can lead to rapid divergence, thereby enhancing long-term stability analyses and population studies in the outer Solar System.

In practice, ML emulators can interpolate between ensembles of numerical migration simulations, enabling rapid exploration of parameter space and Bayesian inference of migration scenarios consistent with observational constraints. Crucially, current studies emphasize that ML models must remain anchored to physically motivated dynamics, as purely data-driven extrapolation is unreliable in chaotic regimes. Hybrid dynamical--statistical frameworks, therefore, represent the most promising avenue for future progress.


\section{Open Problems and Future Directions}
\label{sec:open_problems}

Despite decades of progress in understanding the dynamical structure and evolution of the Kuiper Belt, many fundamental questions remain unresolved. The complexity of the trans-Neptunian region, shaped by resonances, chaotic transport, planetary migration, and physical evolution, continues to challenge both theoretical models and observational interpretation. In this section, we outline key open problems and identify promising directions for future research in Kuiper Belt dynamics.

One of the most persistent challenges concerns the initial conditions of the primordial planetesimal disk. While current models can reproduce many observed features of the Kuiper Belt, such as resonant populations and the cold classical belt, these successes often depend sensitively on assumptions about the disk's radial extent, mass distribution, and initial dynamical temperature. Constraining these parameters remains difficult due to degeneracies between disk properties and migration histories. Progress will likely require tighter integration between planet formation theory, dynamical modeling, and observational constraints on size distributions and compositional gradients.

The migration history of Neptune remains a central, yet incompletely resolved, problem. Although a broad consensus exists that Neptune migrated outward by several astronomical units, the detailed nature of this migration—whether smooth or stochastic, rapid or prolonged—remains debated. In particular, the timing, magnitude, and dynamical consequences of potential semimajor axis jumps during planetary instability are not fully constrained. Future work combining high-precision orbital data with data-driven inference methods holds promise for narrowing the range of viable migration scenarios.

The role of chaos and long-term transport in shaping Kuiper Belt populations also presents unresolved questions. While chaotic diffusion and resonance sticking are well-established mechanisms, quantifying their cumulative effects over the age of the Solar System remains challenging. In particular, the rates at which objects leak from resonances, transition between populations, or enter the Centaur and comet reservoirs are still uncertain. Improved chaos indicators, longer integrations, and statistical approaches will be essential for refining these estimates.

From an observational perspective, survey completeness and bias correction continue to limit our ability to draw definitive conclusions. Although recent surveys have dramatically increased the number of known Kuiper Belt objects, the observed population remains a biased subset of the true distribution. Accurately modeling these biases and incorporating them into dynamical inference frameworks is essential for meaningful comparison between models and data. Upcoming facilities and coordinated survey strategies are expected to substantially improve this situation.

The integration of machine learning and data-driven methods into Kuiper Belt dynamics is still in its early stages. While ML techniques have demonstrated impressive capabilities in surrogate modeling, chaos detection, and parameter inference, their robustness and interpretability remain active areas of research. Ensuring that ML-driven conclusions are physically meaningful and not artifacts of training data or model architecture will require careful validation and continued collaboration between dynamical theorists and data scientists.

Another intriguing direction involves the possible influence of additional massive bodies in the outer Solar System. Hypotheses invoking distant, unseen perturbers have been proposed to explain certain orbital clustering patterns among extreme trans-Neptunian objects. While such ideas remain controversial, they underscore the sensitivity of the Kuiper Belt to weak perturbations and highlight the need for comprehensive dynamical models that consider a wide range of possibilities.

Finally, future spacecraft missions and observational campaigns promise to revolutionize our understanding of the Kuiper Belt. In situ measurements provide invaluable ground truth for interpreting remote observations and testing dynamical models. Continued exploration, combined with advances in numerical methods and data analysis, will likely uncover new structures, populations, and dynamical processes that challenge existing~\mbox{paradigms}.


\section{Conclusions}
\label{sec:conclusions}

The Kuiper Belt represents a dynamically rich and historically informative component of the Solar System, preserving signatures of planet formation, migration, and long-term gravitational evolution. Through a combination of resonant dynamics, chaotic transport, and secular interactions, the trans-Neptunian region records the imprint of Neptune's migration and the broader dynamical history of the outer Solar System.

In this review, I have summarized the current understanding of small-body dynamics in the Kuiper Belt, with an emphasis on mean-motion and secular resonances, stability indicators, and the resulting orbital structure. The distinction between cold classical, hot classical, resonant, scattering, and Centaur populations reflects fundamentally different dynamical pathways, each shaped by specific mechanisms such as resonance sweeping, chaotic diffusion, and planetary encounters. The preservation of dynamically cold populations and kernel structures places particularly strong constraints on models of Neptune's migration.

The role of chaos detection methods has been highlighted (including Lyapunov exponents, MEGNO, SALI/GALI, and Lagrangian descriptors) in revealing the fine structure of the Kuiper Belt phase space and quantifying long-term stability. These tools remain essential for understanding resonance overlap, transport mechanisms, and population transitions, especially in regions where regular and chaotic dynamics coexist.

Recent advances in machine learning and data-driven modeling offer powerful new approaches for accelerating simulations, classifying stability, and reconstructing planetary migration histories. When combined with traditional dynamical methods, these techniques enable efficient exploration of high-dimensional parameter spaces and more rigorous statistical comparison between models and observations. Their continued development and careful integration into dynamic workflows promise to substantially enhance our ability to interpret the growing body of observational data.

Despite significant progress, many open questions remain, including the detailed nature of Neptune's migration, the origin of the cold classical belt, and the long-term effects of chaotic transport and physical evolution. Addressing these challenges will require coordinated efforts across theory, observation, and computation, as well as the continued refinement of both classical and data-driven methodologies.


In conclusion, the study of Kuiper Belt dynamics stands at a crossroads, where classical celestial mechanics, high-performance computing, and data-driven methodologies converge. Addressing the open problems outlined here will require interdisciplinary approaches, sustained observational efforts, and continued methodological innovation. As these challenges are met, the Kuiper Belt will continue to serve as a uniquely informative laboratory for probing the formation and evolution of planetary systems, both within our Solar System and beyond.

\vspace{6pt}

{Funding. This 
work was supported by the Hungarian National Research, Development and Innovation Office, under Grant Nos. NKKP\_AD-153324, NKKP\_AD-152888, and TKP2021-NKTA-64, financed by the Ministry of Culture and Innovation of Hungary.}
\vspace{3mm}

{Data availability. The Julia code for reproducing Figure \ref{fig:giordano2018} in the paper can be found in the following link: \url{https://doi.org/10.5281/zenodo.20799458}.} 
\vspace{3mm}

{Acknowledgments. During 
  the preparation of this manuscript, the author used Anthropic Claude Sonnet 4.6 for the purposes of text and English grammar polishing. The author has reviewed and edited the output and takes full responsibility for the content of this publication. I would also like to thank E. Kőv\'ari and J. Daquin for their kind assistance and for permitting me to use their data and figures in this manuscript. 
}
\vspace{3mm}

{Conflicts of interest. The author declares no conflicts of interest.} 

\appendix
\section{Recurrence Plots and Recurrence Quantification Analysis}
\label{sec:appA}

Recurrence plots (RPs) were introduced \cite{Eckmann1987} as a graphical tool to visualize the recurrence of states in phase space for dynamical systems. The basic idea is that trajectories of deterministic systems repeatedly return arbitrarily close to previous states. Given a time series $\{\mathbf{x}_i\}_{i=1}^{N}$ in a reconstructed phase space, the recurrence matrix is defined as
\begin{equation}
R_{ij}(\varepsilon) = \Theta\big(\varepsilon - \|\mathbf{x}_i - \mathbf{x}_j\|\big),
\end{equation}
where $\|\cdot\|$ denotes a norm in phase space, $\varepsilon$ is a recurrence threshold, and $\Theta$ is the Heaviside function. The matrix $R_{ij}$ is binary and symmetric, and its graphical representation reveals patterns corresponding to periodicity, quasi-periodicity, chaos, or stochasticity. Diagonal line structures indicate deterministic dynamics and predictability, whereas fragmented or short diagonals are typical of chaotic motion \cite{Marwan2007}.

To move beyond visual inspection, Zbilut and Webber \cite{Zbilut1992,Webber1994} developed recurrence quantification analysis (RQA), introducing statistical measures derived from the geometry of the recurrence matrix. Among the most widely used quantities are the recurrence rate (RR), which measures the density of recurrence points; determinism (DET), defined as the fraction of recurrence points forming diagonal lines and thus quantifying predictability; the average diagonal line length $L$, related to the inverse of the largest Lyapunov exponent; and laminarity (LAM), associated with vertical line structures and intermittent behavior. These measures provide quantitative diagnostics of dynamical regimes and transitions between regular and chaotic motion.

Recurrence-based methods have found broad applications in nonlinear dynamics, ranging from climate variability \cite{Breitenbach2010} and neuroscience to engineering \cite{Godavarthi2018} and astrophysics \cite{Kopacek2010,Mohan2025}. In celestial mechanics, recurrence plots have been employed to detect chaos in few-body systems, characterize sticky motion near resonances, and identify dynamical transitions in planetary and small-body systems \cite{Kovacs2020}. Their main advantages are conceptual simplicity, applicability to relatively short time series, and robustness in detecting subtle dynamical changes. As such, recurrence analysis offers a complementary perspective to Lyapunov-based and entropy-based chaos indicators, particularly in weakly chaotic, resonance-dominated regimes such as those encountered in Kuiper Belt dynamics.




\bibliographystyle{plainnat}
\bibliography{references}

\end{document}